\documentclass[twocolumn,amsmath,superscriptaddress,amssymb,nopacs,nofootinbib]{revtex4-2}

\usepackage[usenames,dvips]{color}
\usepackage{graphicx}
\usepackage{dcolumn}
\usepackage{bm}
\usepackage{mathtools}
\usepackage{epstopdf}
\usepackage{esint}
\usepackage{xcolor}
\usepackage{dsfont}
\usepackage{nicefrac}

\usepackage[colorlinks=true,citecolor=magenta,linkcolor=magenta, urlcolor=magenta]{hyperref}

\newcommand{\bea}{\begin{eqnarray}}
\newcommand{\eea}{\end{eqnarray}}
\newcommand{\bt}{\textbf}

\newcommand{\phd}{\phantom{\dag}}
\newcommand{\ph}{\phantom{.}}

\newcommand{\noi}{\noindent}
\newcommand{\no}{\nonumber}

\begin{document}
\def\v#1{{\bf #1}}

\title{Zero-field superconducting vortices and Majorana zero modes pinned\\
by magnetic islands in correlated Rashba systems}

\author{Panagiotis Kotetes}
\email{kotetes@baqis.ac.cn}
\affiliation{Beijing Academy of Quantum Information Sciences, Beijing 100193, China}

\author{Brian M. Andersen}
\email{bma@nbi.ku.dk}
\affiliation{Niels Bohr Institute, University of Copenhagen, DK-2200 Copenhagen, Denmark}

\vskip 1cm

\begin{abstract}
We propose a route for pinning zero-field superconducting vortices in systems which
are exchange-coupled to magnetic islands and feature Rashba spin-orbit coupling. We
consider islands with sizes which greatly exceed those of the vortex cores and possess out-of-plane magnetic
moments. A crucial ingredient of our approach is that it considers superconductors which are governed by magnetic
correlations without, however, exhibiting long range magnetic order. The arising total magnetization
is inhomogeneous and its gradients generate a nonzero vorticity in the superconducting phase.
Vortices become energetically stable due to the energy reduction brought about from the generation
of electronic magnetization. Using our developed framework, we make concrete predictions for the emergence of zero-field vortices and Majorana zero modes in superconducting topological insulator surfaces and planar Rashba superconductors. Our theory uncovers a nonstandard path for trapping composite vortex-Majorana excitations in systems which appear to be within experimental reach. \end{abstract}

\maketitle

\section{Introduction} 

The study of superconducting vortices has recently attracted renewed interest since - in certain cases - these
can trap Majorana zero modes (MZMs)~\cite{ReadGreen,Volovik99,FuKane,SauPRL,Hosur,ChiuPRB,Biswas}. MZMs define exotic quasiparticles which are electrically neutral~\cite{
Alicea,CarloRev,Leijnse,Aguado,KotetesTSCchapter}, satisfy non-Abelian exchange statistics~\cite{Ivanov2001} and, thus, appear as unique components for implemen\-ting topologically-protected fault tolerant quantum computing~\cite{Ivanov2001,Kitaev2003,Nayak2008,Alicea2011}. Presently, the to\-po\-lo\-gical superconductors (SCs) FeTe$_{0.55}$Se$_{0.45}$~\cite{TopoSurfaceStatesFeSe1,TopoSurfaceStatesFeSe2} and LiFeAs~\cite{PZhangNatPhys,SongtianPRB,LiFeAsMZMTunable} appear as prominent candidates for experimentally rea\-lizing a variety of vortex-MZM scenarios~\cite{FeTeSeTopo1,FeTeSeTopo2,KonigMZM,weakVMZMsQin,CKChiu,Pathak,XinLiu}. A number
of experimental evidences supporting this possibility have
been provided in the case of vortices stabilized by
an external magnetic field~\cite{hongding1,hongding2,LiFeSeMZM,Machida,Lingyuan,LingyuanZhu}. However, there also
exist experiments suggesting that MZMs emerge in these
compounds even in the absence of an external field~\cite{JiaXinYin}. One possibility is that the ari\-sing MZMs get pinned by clu\-sters of interstitial Fe atoms, which have already been detected in these systems~\cite{Thampy}. Signatures asso\-cia\-ted with defect-pinned MZMs have also been experimentally provided in magnetic adatom islands deposited on top of Pb~\cite{Cren}. MZMs also appear here in the absence of an external magnetic field, and possible explanations include the pre\-sen\-ce of vortices in the spin-orbit coupling (SOC) field~\cite{Cren} or magnetic texture defects~\cite{Cren,NakosaiBalatsky,KlinovajaSkyrmion,Mirlin,Garnier,Mesaros}. Note that the involvement of Abrikosov vortices was fully dismissed in the experiment of Ref.~\onlinecite{Cren}, since these are expected to extend over an area determined by the superconducting coherence length which, in these hybrids, exceeds by far the radius of the magnetic island. More recently, the possibility of zero-field vortices has also been discussed in the context of 4Hb-TaS$_2$, where a large magnetization was detected when entering the superconducting state~\cite{Persky2022}. A possible theoretical explanation for this effect has been analyzed in terms of spontaneous vortices generated by charged magnetic inclusions with magnetoelectric supercurrent contributions generated due to Rashba SOC~\cite{Levitan2025}.

The above experiments call for identifying alternative routes to stabilize superconducting vortices without residing to the application of a magnetic field. Prior works in this direction have considered the possibility of pinning vortices by spin-to-flux conversion which relies on the Zeeman effect~\cite{VarmaSpontaneous,Tachiki1979,Tachiki1980,Kuper,TewariSpontaneous}. Such a situation naturally takes place in ferromagnetic SCs and experimental signatures of so-called spontaneous vortices have been already captured in certain materials~\cite{NgSpontaneous,DeguchiSpontaneous,WHJiao}. A similar phenomenon also emerges in ferromagnet-SC interfaces~\cite{BuzdinFM}, where the required flux for stabilizing vortices is provided by the stray field of a ferromagnetic insulator.  Additional pathways to stabilize vortices open up when the Zeeman coupling is neither the only nor the primary mecha\-nism. A prominent means to induce a nonzero flux relies on the presence of a nonzero magnetoelectric coupling, which appears when inversion symmetry is broken. Here, one finds physical realizations in which a superconducting vortex is stabilized in ferromagnet-SC interfaces due to the pre\-sen\-ce of Rashba SOC~\cite{BuzdinSOC}, or, cases in which the currents generated by skyrmions stabilize a nonzero vorticity for the superconducting phase~\cite{Hals,BuzdinNeel}. Aside from the above possibilities, a more recent theo\-retical work showed that a vortex can be induced on a to\-po\-lo\-gi\-cal insulator (TI) surface by a magnetic impurity~\cite{KunJiang}. For this to happen, the impurity needs to be coupled to the total angular momentum of the TI electrons, which is achieved via combined magnetic exchange and local Elliot-Yafet type of couplings. This is in fact a pre\-re\-qui\-si\-te for rendering the vortex phase as the most stable ground state solution, since this suppresses the ener\-gy contribution of the Caroli - de Gennes - Matricon (CdGM) vortex core states~\cite{CdGM}. Notably, recent measurements have provided further evi\-den\-ce for the experimental realization of the associated so-called quantum anomalous vortex~\cite{QAHVexpEarly,QAHVexp}.

\begin{figure*}[t!]
\begin{center}
\includegraphics[width=0.76\textwidth]{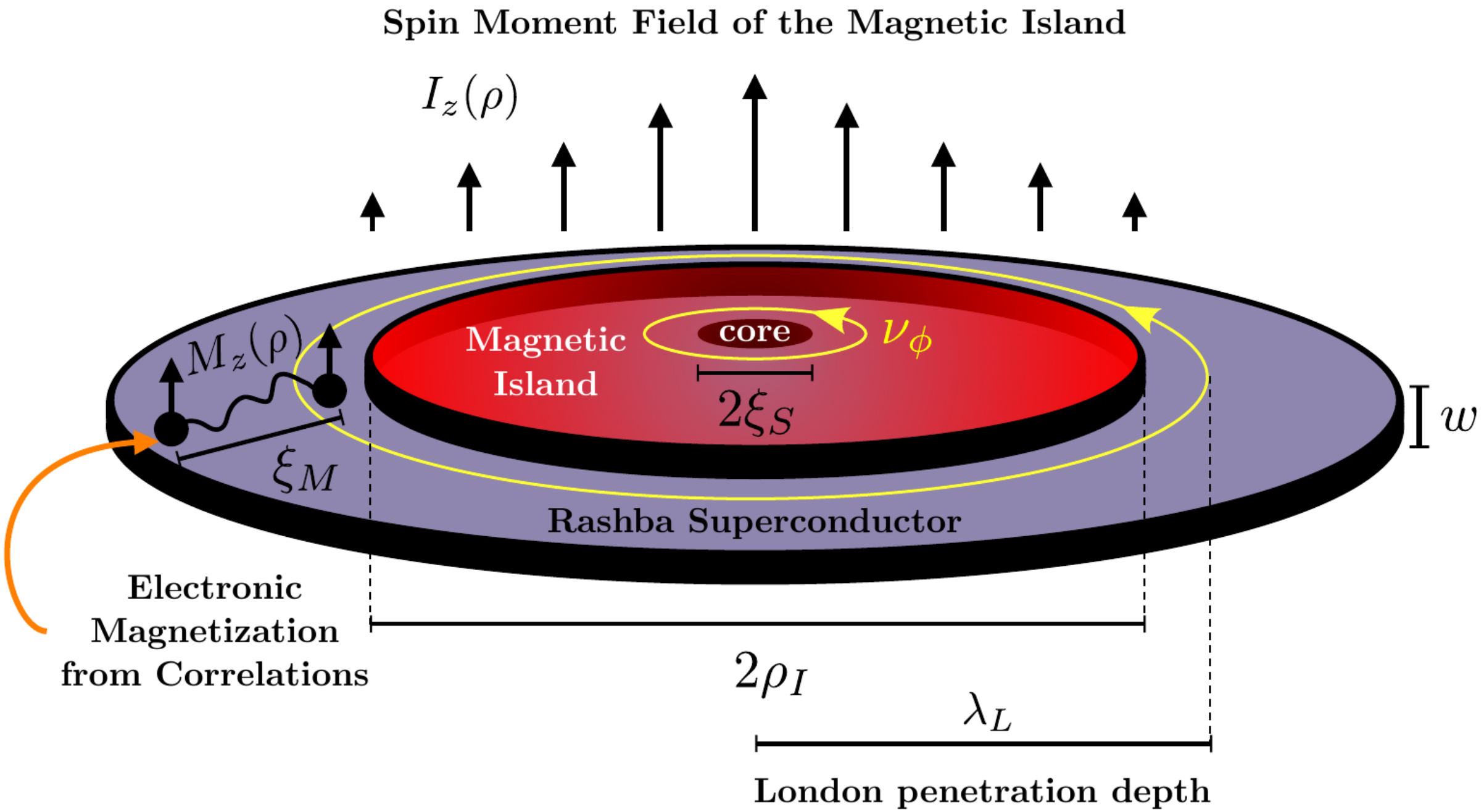}
\end{center}
\caption{Illustration of an extended magnetic impurity, i.e., a magnetic island, which is embedded in a quasi-2D SC of thickness $w$, and is dictated by a Rashba-type SOC. We consider a type-II conventional spin-singlet SC which can harbor superconducting vortices. This implies that its London penetration depth $\lambda_L$ exceeds the superconducting coherence length $\xi_S$. The radius $\rho_I$ of the magnetic island is assumed to be much larger than $\xi_S$. In addition, the spin moment of the magnetic island, which is depicted with black arrows, is assumed to be polarized out of the plane of the Rashba superconducting host. By means of the Zeeman effect and the magnetoelectricity mediated by the Rashba SOC, the spin moment of the island becomes converted into a magnetic flux which induces a vortex with vorticity $\nu_\phi$, under conditions that we specify in this work. Key aspect of our approach is that we also account for possible magnetic correlations that lead to the appearance of electronic magnetization in the SC upon adding the magnetic island. The properties of the induced magnetization are controlled, among others, by the magnetic correlation length $\xi_M$ which is also much larger than $\xi_S$. The spin moment of the island is only exchange-coupled to the electrons of the substrate SC, and is assumed to be sufficiently extended to not induce any Yu-Shiba-Rusinov states~\cite{YSR,Heimes,CommentOnYSR}.}
\label{fig:Figure1}
\end{figure*}

In this Manuscript, we propose an alternative mechanism for stabilizing superconducting vortices and MZMs in generic two-dimensional (2D) Rashba SCs in the absence of externally applied magnetic fields. Within our theoretical proposal, a vortex is pinned by an extended magnetic impurity, i.e., a magnetic island, which is solely exchanged-coupled to the 2D electrons of the SC, see the schematic depiction in Fig.~\ref{fig:Figure1}. The spin moment of the island is assumed to be orien\-ted out-of-the plane due to an easy-axis spin anisotropy imposed by crystal fields. Zeeman and Rashba magnetoelectric effects convert the spin moment of the island into magnetic flux which, in turn, sources a nonzero vorticity in the superconducting phase of the SC. A unique feature of our theoretical approach is the assumption of non-negligible magnetic correlations in the Rashba SC. While these correlations are assumed not to establish  long-ranged magnetic order, they yet remain of great importance, since they mediate the induction of electronic magnetization in the SC. The induced magnetization ``dresses'' the spin moment of the magnetic island, an effect which in certain cases is found to substantially favor the pinning of superconducting vortices with large vorticity values. This is because the electrons of the SC can now exploit their coupling to the magnetic island to further reduce the ener\-gy of the entire system.

Noteworthy, a magnetic-correlations-mediated vortex scenario may be suitable for FeTe$_{0.55}$Se$_{0.45}$, which is known to be in the vi\-ci\-ni\-ty of a magnetic instabi\-li\-ty~\cite{FeTeSeMagnetic,Kreisel_review}. Nonetheless, the potential relevance of our proposal is neither restricted to FeTe$_{0.55}$Se$_{0.45}$ nor TI materials, but is instead applicable to a wider class of 2D Rashba SCs which may be even to\-po\-lo\-gi\-cal\-ly trivial, but exhibit magnetic correlations. To showcase the breadth of the applicability of our proposal, we first study the vortex state and its energy stability based on a phe\-no\-me\-no\-lo\-gi\-cal Ginzburg-Landau (GL) theory which describes the interplay between the electronic magnetization and the electromagnetic vector potential. We elaborate on the various distinct mechanisms which me\-dia\-te the spin-to-vorticity conversion and enable the vortex formation.

By con\-si\-de\-ring a concrete exemplary profile for the spatial distribution of the spin moment carried by the magnetic island, we provide exact analytical expressions for the spatial profiles of the magnetization and the vector potential which develop in the vortex ground state. Moreover, we identify the vorticity value of the superconducting phase which is stabilized upon varying relevant physical quantities, such as, the total spin moment carried by the magnetic island, the radius of the magnetic island, the magnetic correlation length, and the London penetration depth of the SC. Having explored our zero-field vortex pinning mechanism using a general framework, we subsequently apply it to two concrete platforms, i.e., superconducting TI surface states and Rashba SCs. By employing representative values for the various ensuing microscopic parameters, we identify which requirements need to be satisfied so that our mechanism applies to the two types of systems. Finally, we conclude by discussing the properties of the MZMs that can, in turn, become bound to the magnetic-island-pinned vortices.

The manuscript is organized in the  following fashion. In Sec.~\ref{sec:Microscopic} we introduce a representative model Hamiltonian for the type of systems we are considering. This section sets the stage for the GL approach to be employed. Afterwards, in Sec.~\ref{sec:GL0}, we introduce the main assumptions of this work and set up the GL method in the case of negligible magnetic correlations. In this section, we identify the spatial profile of the magnetic field and the superconducting phase by assuming a concrete spatial profile for the spin moment carried by the magnetic island. In addition, we evaluate the vortex ground state energy, and explore the phase winding induced by the magnetic island upon varying the parameters of the problem. In Sec.~\ref{sec:GL} we carry out a similar program, with the crucial dif\-fe\-ren\-ce that we consider the possi\-bi\-li\-ty of non-negligible magnetic correlations which lead to the emergence of electronic magnetization. In Sec.~\ref{sec:RepresentativeModels}, we go beyond our general GL approach and its purely phenomenological exploration, and consider two concrete systems. These concern superconducting topological surface states and Rashba metals. As a follow-up of this analysis, Sec.~\ref{sec:VortexMZMs} discusses the properties of the arising MZMs in each one of these two representative platforms. The conclusions of our investigation are presented in Sec.~\ref{sec:Conclusions}, while further supportive materials and technical details are given in the accompanying Appendices~\ref{app:AppendixA}-\ref{app:AppendixMZM}.

\section{Representative Microscopic Model}\label{sec:Microscopic}

In order to facilitate the presentation of our GL ana\-ly\-sis and prepare the reader for the subsequent investigation of the two concrete systems of interest, we introduce a model Hamiltonian which is representative of the time-reversal-symmetric SCs of interest here. Unless otherwise stated, in this work we mainly consider homogeneous 2D Rashba SCs under the influence of: (i) the spin moment background field $I_z(\bm{r})$ generated by the magnetic island, (ii) the out-of-plane magnetization $M_z(\bm{r})$ that is to be self-consistently determined, and (iii) the in-plane components $A_{x,y}(\bm{r})$ of the vector potential which ge\-ne\-ra\-te the out-of-plane magnetic induction $B_z(\bm{r})$. At this point we remark that throughout this work the spin-moment and magnetization fields, i.e., $I_z(\bm{r})$  and $M_z(\bm{r})$, are always expressed in ener\-gy units. We also note that throughout our manuscript $\bm{r}=(x,y)$ denotes the position vector in 2D space, since all the fields are considered to be independent of the out-of-plane $z$ coordinate.

Such systems are mo\-de\-led here using the Hamiltonian:
\begin{align}
{\cal H}=\frac{1}{2}\int d\bm{r}\,\bm{\Psi}^\dag(\bm{r})\hat{\cal H}(\bm{r})\bm{\Psi}(\bm{r}),
\label{eq:Hamiltonian}
\end{align}

\noi with the Bogoliubov - de Gennes (BdG) Hamiltonian matrix given by:
\begin{align}
\hat{\cal H}(\bm{r})=\hat{{\cal H}}_0(\hat{\bm{\pi}})-\left\{I_z(\bm{r})+M_z(\bm{r})-\frac{g\mu_B}{2}\big[\bm{\nabla}\times\bm{a}(\bm{r})\big]_z\right\}\sigma_z,\label{eq:BdGHamiltonian}
\end{align}

\noi which is defined in the basis set by the spinor $\bm{\Psi}^{\dag}(\bm{r})=\big(\psi_\uparrow^\dag(\bm{r}),\,\psi_\downarrow^\dag(\bm{r}),\,\psi_\downarrow(\bm{r}),\,-\psi_\uparrow(\bm{r})\big)$. Here, $\psi_\sigma(\bm{r})/\psi_\sigma^\dag(\bm{r})$ is an operator that annihilates/creates an electron at position $\bm{r}$ with spin projection $\sigma=\uparrow,\downarrow$. Throughout this work, any Hamiltonian matrix, such as the above, is expressed with Kronecker pro\-ducts which are constructed using the Pauli matrices $\bm{\tau}$ and $\bm{\sigma}$, along with their respective unit matrices $\mathds{1}_{\tau,\sigma}$. These are defined in Nambu and spin spaces, respectively. We will omit writing the Kronecker product symbol $\otimes$ and unit matrices in the remainder.

The BdG Hamiltonian is expressed in a compact fa\-shion by employing the gauge invariant momentum ope\-ra\-tor $\hat{\bm{\pi}}=\hat{\bm{p}}+e\tau_z\bm{a}(\bm{r})$, which is expressed in terms of the gauge invariant vector potential:
\begin{align}
\bm{a}(\bm{r})=\bm{A}(\bm{r})+\Phi_0\frac{\bm{\nabla}\phi(\bm{r})}{2\pi},
\end{align}

\noi where $e>0$ denotes the electric charge unit, $\hat{\bm{p}}=-i\hbar\bm{\nabla}$ corresponds to the momentum operator, $\hbar$ is the reduced Planck constant, $\Phi_0=h/2e$ defines the superconducting flux quantum, and $\phi(\bm{r})$ is the spatially-varying superconducting phase field. In Eq.~\eqref{eq:BdGHamiltonian}, $g$ denotes the Land\'e gyromagnetic factor and $\mu_B$ the Bohr magneton. Note that, here, the Zeeman term includes the curl of the gauge invariant vector potential $\bm{a}(\bm{r})$ instead of $\bm{A}(\bm{r})$~\cite{CommentOnZeeman}.

In the BdG Hamiltonian given in Eq.~\eqref{eq:BdGHamiltonian}, we have also inserted the bare Hamiltonian, i.e., the one which is defined in the absence of the magnetic island and reads as:
\begin{align}
\hat{\cal H}_0(\hat{\bm{p}})=\tau_z\left[\frac{\hat{\bm{p}}^2}{2m}-\mu+\upsilon_R\big(\hat{p}_x\sigma_y-\hat{p}_y\sigma_x\big)\right]+\Delta\tau_x.\label{eq:H0}
\end{align}

In the above, the variable $\upsilon_R>0$ controls the strength of the Rashba SOC, $m>0$ defines the effective mass of the electrons, and $\mu$ is the chemical potential. The modulus of the pairing gap $\Delta\geq0$ is not treated self-consistently. Nonetheless, the presence of the composite island-vortex defect is accounted for by assuming that $\Delta$ vanishes in the vortex core, whose radius is determined by the superconducting coherence length $\xi_S$. We discuss possible limitations of our non-self-consistent approach at a later stage. We addi\-tio\-nal\-ly note that the model allows us to smoothly interpolate between the case of a Rashba metal and the surface states of a three-dimensional (3D) TI. In fact, to simulate the topological surface states, we consider the effective mass to be sufficiently large, so that it does not lead to two helical branches within the energy cutoff which sets the regime of validity of our model~\cite{Anatomy}.

\section{Vortex Ground State in the Absence of Magnetic Correlations}\label{sec:GL0}

We first explore the emergence of a supercon\-duc\-ting vortex when the electrons of the SC are coupled to the spin moments induced by a magnetic island, without accounting for the feedback of the electronic magnetization to the formation of the vortex. Under this assumption, we now investigate the vortex solution and its stability.

\subsection{Ginzburg-Landau Theory} \label{sec:GL0functional}

We first discuss the nucleation of our nonstandard vortices in a generic fa\-shion through a phenomenological GL approach. Within this framework, we  assume that the superconducting coherence length $\xi_S$ is much smaller than all the other lengthscales entering the problem, such as, the London penetration depth $\lambda_L$, the size of the island which is set by a cha\-rac\-te\-ri\-stic lengthscale $\rho_I$, and the correlation length $\xi_M$ gover\-ning the magnetic correlations. The latter lengthscale will become relevant later on, when we include the magnetic interactions. Since the size of the vortex core is con\-si\-de\-red negligible, we can focus on the physics outside the core, as it is also cu\-sto\-ma\-ry for magnetic-field-induced vortices~\cite{deGennes}. In fact, it is exactly the assumption of a negligible vortex core that further allows us to consider that the superfluid density is spatially uniform outside the core, and obtains a value which is determined by the pairing gap $\Delta$ of the bulk SC.

Given the above considerations, we can express the GL functional solely in terms of the spatial spin profile of the magnetic island which is described by the classical background field $I_z(\bm{r})$, the magnetic induction $B_z(\bm{r})=\big[\bm{\nabla}\times\bm{A}(\bm{r})\big]_z$, the electromagnetic vector potential $\bm{A}(\bm{r})$, and the superconducting phase $\phi(\bm{r})$. Notably, in our proposal, $I_z(\bm{r})$ acts as a source field for the magnetic induction $B_z(\bm{r})$. The latter couples to $I_z(\bm{r})$ due to the Zeeman effect, as well as due to magnetoelectric effects arising from the presence of the Rashba SOC.

The GL functional in the absence of magnetic correlations is obtained by expanding the ener\-gy of the system up to first order with respect to the background field $I_z(\bm{r})$. By doing this, we obtain the following expression for the 3D GL energy density:
\bea
E_{\rm GL}^{{\rm 3D}}(\bm{r},z)&=&\frac{B_z^2(\bm{r})}{2\mu_0}+\frac{D}{w}\frac{\bm{a}^2(\bm{r})}{2}+\frac{{\cal X}}{w}\frac{\bm{a}(\bm{r})\cdot\big[\bm{\nabla}\times\hat{\bm{z}} I_z(\bm{r})\big]}{2}\no\\
&&+\frac{{\cal X}}{w}\frac{I_z(\bm{r})\hat{\bm{z}}\cdot\big[\bm{\nabla}\times\bm{a}(\bm{r})\big]}{2}\,,
\label{eq:GL0Bulk}
\eea

\noi where $\hat{\bm{z}}$ defines the unit vector in the out-of-plane $z$ direction of the planar SC. Throughout this work, the gra\-dient $\bm{\nabla}$ is restricted to the in-plane gra\-dient vector $(\partial_x,\partial_y)$.

We remark that, aside from accounting for the Zeeman and Rashba magnetoelectric couplings to the background field $I_z(\bm{r})$, here we discard any other possible modifications to the part describing the electromagnetic field. This is because in the situations discussed in this work, the vortex formation is not driven by an external magnetic field, but instead results from the effective flux ge\-ne\-ra\-ted by the spin moment field of the magnetic island. Moreover, we remark that the coupling between $\bm{a}(\bm{r})$ and $I_z(\bm{r})$ enters in a symmetrized fa\-shion. This ensures that no ambiguous total derivative terms appear, which would otherwise contribute to the surface energy~\cite{CommentOnSymmetrization}.

We now proceed with detailing the nature and role of the various coefficients and terms appearing in Eq.~\eqref{eq:GL0Bulk}. First of all, $\mu_0$ and $D$ denote the vacuum permeability and the superfluid stiffness, which are the two quantities that govern the electromagnetic properties of the Rashba SC in the absence of the magnetic island. The re\-mai\-ning terms describe the coupling of the spin moment field of the magnetic island to the gauge inva\-riant vector potential. This coupling is mediated by Zeeman and Rashba magnetoelectric effects, with coefficients $\chi_Z$ and $\chi_R$, respectively. These comprise the total coupling ${\cal X}=\chi_Z+\chi_R$. The strength of the Zeeman-effect-mediated coupling is defined as~\cite{Anatomy}:
\begin{align}
\chi_Z=\frac{1}{2}g\mu_B\chi_\perp^{\rm spin}\,,\label{eq:ZeemanConversion}
\end{align} 

\noi where $\chi_\perp^{\rm spin}$ corresponds to the out-of-plane spin su\-sceptibility which is positive/negative for a paramagnetic/diamagnetic SC. In addition, ana\-ly\-tical expressions have been derived for $\chi_R$ in Refs.~\onlinecite{PershogubaCurrents,Anatomy} for a Rashba metal in both the normal and superconducting phases, and in Ref.~\onlinecite{Anatomy} for a TI surface~\cite{CommentOnSuc}, once again, in both normal and superconducting phases. Notably, when the magnetoelectric coupling $\chi_R$ is nonzero, it ena\-bles the appearance of a superconducting diode effect from magnetization gradients, which was recently proposed in Ref.~\onlinecite{Anatomy} and analyzed in Refs.~\onlinecite{Roig,KotetesDiode}.

The GL functional introduced in Eq.~\eqref{eq:GL0Bulk} is defined in three spatial dimensions and the total energy of the system is given by integrating out the above energy density over the infinite $xy$ plane and the thickness $w$ of the sample, i.e., we have $E_{\rm GL}^{\rm 3D}=\int d\bm{r}\int_0^wdz\,E_{\rm GL}^{\rm 3D}(\bm{r},z)$. Throughout this work, the sample thickness $w$ is considered to be sufficiently small to allow for the electronic degrees of freedom to effectively exhibit a strictly 2D behavior. To facilitate the notation of the upcoming analysis, in the remainder we explore the properties of an effective GL functional $E_{\rm GL}(\bm{r})$ defined in two spatial dimensions, obtained through the relation $E_{\rm GL}(\bm{r})=\int_0^wdz\,E_{\rm GL}^{{\rm 3D}}(\bm{r},z)$. Furthermore, we choose a unit sy\-stem that allows us to write the expression for the GL functional in the following simpler and more compact form:
\bea
E_{\rm GL}(\bm{r})&=&\frac{B_z^2(\bm{r})}{2}+\frac{\bm{a}^2(\bm{r})}{2\lambda_L^2}+\varGamma\frac{\bm{a}(\bm{r})\cdot\big[\bm{\nabla}\times\hat{\bm{z}} I_z(\bm{r})\big]}{2}\no\\
&&+\varGamma\frac{I_z(\bm{r})\hat{\bm{z}}\cdot\big[\bm{\nabla}\times\bm{a}(\bm{r})\big]}{2}\,.
\label{eq:GL0}
\eea

\noi In this unit system, the gauge invariant vector potential becomes rescaled according to $\bm{a}(\bm{r})\mapsto\sqrt{\mu_0/w}\,\bm{a}(\bm{r})$, thus equivalently implying that the various coefficients
are replaced by new ones given by the formulas:\begin{align}
\varGamma=\sqrt{\frac{\mu_0}{w}}\,{\cal X},\,\phd\gamma_{R,Z}=\sqrt{\frac{\mu_0}{w}}\,\chi_{R,Z},\,\phd{\rm and}\,\phd \lambda_L=\sqrt{\frac{w}{\mu_0D}}\,.   
\end{align}

\subsection{Little-Parks vs Meissner effect} 
\label{sec:LittleParks}

Before procee\-ding with the study of the vortex state, it is important to discuss the two most prominent regimes which control the vortex physics in the present case. For this purpose, we infer the electric current density $\bm{J}(\bm{r})$ in the absence of magnetic fluctuations, which is given by the expression:
\begin{align}
\bm{J}(\bm{r})=-\varGamma\bm{\nabla}\times\hat{\bm{z}}I_z(\bm{r})-\bm{a}(\bm{r})/\lambda_L^2\,.
\label{eq:Current0}
\end{align}

\noi We first note that the second term in Eq.~\eqref{eq:Current0} is associated with the quantization of the fluxoid in the case where Meissner screening currents are substantial. Spe\-ci\-fi\-cal\-ly, by considering a path ${\cal C}=\partial{\cal S}$ inside the supercon\-duc\-ting region which encircles an area ${\cal S}$ beyond which the current is zero, one finds from Eq.~\eqref{eq:Current0}:
\bea
\oiint_{{\cal S}}d{\cal S}\left[B_z(\bm{r})-\varGamma\big(\lambda_L\bm{\nabla}\big)^2I_z(\bm{r})\right]=\nu_\phi\Phi_0,
\label{eq:FluxoidQuantization}
\eea

\noi where $\nu_\phi\in\mathbb{Z}$ defines the vorticity of the superconducting phase field through the relation $\ointctrclockwise_{\cal C}d\bm{r}\cdot\bm{\nabla}\phi(\bm{r})=-2\pi\nu_\phi$. 

For a number of the cases of interest, however, $\lambda_{L}$ is assumed to be much larger than the lengthscales dictating the vanishing of the ma\-gne\-ti\-za\-tion. For instance, in the case of FeTe$_{0.55}$Se$_{0.45}$ the London penetration depth for out-of-plane magnetic fields is $\lambda_{L}\sim1500\,{\rm nm}$~\cite{Toulemonde}. Hence, in such an event, the diamagnetic part of the current, i.e., the one $\propto\bm{A}$, which is responsible for the Meissner screening of the magnetic field, has a subleading contribution to the total electric current density in the vi\-ci\-ni\-ty of the vortex. In fact, in the majority of situations which are relevant for our study, the Meissner scree\-ning currents become negligible, thus allowing for the flux induced by the magnetic island to pe\-ne\-tra\-te substantially inside the sample. Hence, the value of vorticity $\nu_\phi$ will have to adjust itself in order to minimize the ener\-gy of the system given the induced magnetic flux, in analogy to the situation taking place in the Little-Parks effect~\cite{LittleParks}. Therefore, in complete analogy to the Little-Parks effect, Eq.~\eqref{eq:Current0} implies that a nonzero current will circulate in the sample due to the ineffective Meis\-sner screening.

\subsection{Vortex Solution}
\label{sec:EOM}

By exploiting the Maxwell equation $\bm{\nabla}\times\bm{B}(\bm{r})=\bm{J}(\bm{r})$, which is expressed in accordance with the unit system employed here, we find the fundamental relation:
\begin{align}
-\lambda_L^2\bm{\nabla}\times\hat{\bm{z}}H_z(\bm{r})=\bm{a}(\bm{r})\,,\label{eq:curlHz0}
\end{align}

\noi where we introduced the magnetic field:
\begin{align}
H_z(\bm{r})=B_z(\bm{r})+\varGamma I_z(\bm{r})\,.
\end{align}

\noi By further acting with $\bm{\nabla}\times$ on Eq.~\eqref{eq:curlHz0}, we find the equation of motion (EOM):
\begin{align}
\left[1-\big(\lambda_L\bm{\nabla}\big)^2\right]H_z(\bm{r})=\varGamma I_z(\bm{r})+\nu_\phi\Phi_0\delta(\bm{r})\,.
\label{eq:HzFull0}
\end{align}

\noi The above is obtained for $\phi(\bm{r})=-\nu_\phi\theta$ where $\tan\theta=y/x$, which also implies the relation $\hat{\bm{z}}\cdot\big(\bm{\nabla}\times\bm{\nabla}\big)\phi(\bm{r})=-2\pi\nu_\phi\delta(\bm{r})$. Here, it is important to remark that the term $\nu_\phi\Phi_0\delta(\bm{r})$ drops out from Eq.~\eqref{eq:HzFull0} in the regime outside the vortex core. Nonetheless, as it is customary for a vortex core with a negligible size~\cite{deGennes}, we obtain the solution outside the vortex core by first fin\-ding the expression for $H_z(\bm{r})$ using Eq.~\eqref{eq:HzFull0} everywhere in $\rho\in[0,\infty)$, and subsequently by restric\-ting the defining radial coordinate domain to $\rho\in[\xi_S,\infty)$, with the lower bound being set by the superconducting coherence length $\xi_S$.

The solution for the magnetic field $H_z(\bm{r})$ is obtained by inverting the respective 2D Fourier transform $H_z(\bm{q})$ which is given by the definition:
\begin{align}H_z(\bm{r})=\int \frac{d\bm{q}}{(2\pi)^2}\ph e^{i\bm{q}\cdot\bm{r}}H_z(\bm{q}),
\end{align}

\noi with $\bm{q}=(q_x,q_y)$ and $q=|\bm{q}|$. Equation~\eqref{eq:HzFull0} provides the following expression:
\begin{align}
H_z(\bm{q})=\frac{1}{\lambda_L^2}\frac{\varGamma I_z(\bm{q})+\nu_\phi\Phi_0}{q^2+1/\lambda_L^2}\,,\ph\quad
\label{eq:HzFourier0}
\end{align}

\noi where we introduced the Fourier transform:
\begin{align}I_z(\bm{q})=\int d\bm{r}\,e^{-i\bm{q}\cdot\bm{r}}I_z(\bm{r})\,.
\end{align}

When the magnetic island is not present, the Fourier transform yields the standard expression for the magnetic field~\cite{deGennes}, i.e., $H_z(\bm{r})=\nu_\phi H_0K_0(\rho/\lambda_L)$. Here, we have introduced $\rho=\sqrt{x^2+y^2}$ and $H_0=\Phi_0/(2\pi\lambda_L^2)$, while $K_0(z)$ denotes the zeroth order mo\-di\-fied Bessel function of the second kind with $z\in[0,\infty)$. The function $K_0(z)$ bears si\-mi\-la\-ri\-ties to an exponential function decaying away from the origin of the coordinate system. Based on this observation, we also find that the inversion of the Fourier transform of the magnetic field is facilitated by considering the following expression for the spin moment of the magnetic island:
\begin{align}
I_z(\bm{r})=\frac{S_z}{2\pi\rho_I^2}K_0(\rho/\rho_I)\,,\label{eq:SpinMomentProfile}
\end{align}

\noi where $\rho_I$ encodes the spatial extension of the magnetic island and $S_z$ its spin moment. Employing this spatial profile simplifies the ana\-ly\-ti\-cal evalua\-tions since its Fourier transform $I_z(\bm{q})$ is essentially determined by the zeroth order Hankel transform of $I_z(\rho)$. Specifically, we have: 
\begin{align}
I_z(\bm{q})=2\pi\int_0^\infty I_z(\rho)J_0(q\rho)\rho d\rho=\frac{S_z}{\rho_I^2}\frac{1}{q^2+1/\rho_I^2}\,,
\label{eq:HankelK}
\end{align}

\noi where $J_0(z)$ denotes the zeroth order Bessel function of the first kind with $z\in[0,\infty)$. Interestingly, despite the fact that $K_0(z)$ diverges logarithmically as $z\rightarrow0$, integrals involving $K_0(\rho)$ are well behaved. For instance, we have the relations $\int_0^\infty K_0(z)zdz=\int_0^\infty 2K_0^2(z)z dz=1$. Equation~\eqref{eq:HankelK} also implies that the parameter $S_z$ yields the total spin moment of the magnetic island. Here, $S_z/\rho_I^2$ is in energy units and is defined through the relation $S_z=I_z(\bm{q}=\bm{0})=\int d\bm{r}\, I_z(\bm{r})$. We remark that even though we choose a concrete profile for the spin-moment field in order to facilitate the analytical treatment, our conclusions are also qualitatively valid for other types of spatial profiles, as long as these decay away from the vortex core within a characteristic length scale $\rho_I$ and carry a total spin-moment $S_z$. Indeed, our theo\-ry is applicable to more general and irregular spatial profiles, given that any additional spatial variations that these may exhibit -- on top of the principal decaying envelope function with $\rho_I$ -- evolve in space faster than $\rho_I$ and slower than the Fermi wavelength~\cite{CommentOnYSR}, thus leaving $\rho_I$ to be the crucial parameter controlling the spatial profile of the magnetic island. Details for a disk-like profile are given in Ref.~\onlinecite{CommentOnDiskProfile}.

We now proceed with the inversion of the Fourier transform given the spatial profile exhibited by Eq.~\eqref{eq:SpinMomentProfile}. The calculation is straightforward and yields the expressions:
\bea
H_z(\bm{r})&=&H_0\left[\nu_\phi+\frac{\nu_I}{1-(\rho_I/\lambda_L)^2}\right] K_0(\rho/\lambda_L)\no\\
&-&H_0\frac{\nu_I}{1-(\rho_I/\lambda_L)^2} K_0(\rho/\rho_I)\,,\\
B_z(\bm{r})&=&H_0\left[\nu_\phi+\frac{\nu_I}{1-(\rho_I/\lambda_L)^2}\right] K_0(\rho/\lambda_L)\no\\
&-&H_0\left(\frac{\lambda_L}{\rho_I}\right)^2\frac{\nu_I}{1-(\rho_I/\lambda_L)^2} K_0(\rho/\rho_I)\,,
\label{eq:HzProfile0}
\eea

\noi where $\nu_I=\varGamma S_z/\Phi_0$ determines the number of flux quanta which are generated from the total spin moment $S_z$ carried by the magnetic island. We remark that $\nu_I$ is a dimensionless number since in this unit system $\Phi_0$ is not expressed in units of flux. Indeed, in the original SI unit system, where $\Phi_0$ is expressed in units of flux, we have $\nu_I=\mu_0S_z{\cal X}/w\Phi_0$. Since ${\cal X}$ is expressed in units of inverse flux, one indeed confirms that $\nu_I$ is dimensionless.

\subsection{Vortex Stability and Phase Winding}

With the expressions of the magnetic field and induction at hand, we now infer the value of the vorticity $\nu_\phi$ which becomes stabilized in the presence of the background field $I_z(\bm{r})$. For this purpose, we obtain the ener\-gy for the respective vortex ground state. By employing the relations in Eqs.~\eqref{eq:curlHz0} and~\eqref{eq:HzFull0}, which are satisfied by the magnetic field and induction at the vortex ground state extremum of the GL functional, we find the expression:
\begin{align}
E_{\rm vortex}(\bm{r})=\frac{\lambda_L^2}{2}\bm{\nabla}\cdot\big[B_z(\bm{r})\bm{\nabla}H_z(\bm{r})\big]+\frac{1}{2}\varGamma I_z(\bm{r})B_z(\bm{r})\,.
\label{eq:VortexEnergy0}
\end{align}

\noi To obtain the above, we focus on the region outside the vortex core, in which case the Dirac delta function drops out of Eq.~\eqref{eq:HzFull0}. Let us now comment on the structure of the above result. The first term constitutes the vortex surface tension density, accordingly mo\-di\-fied due to the presence of the magnetic island. Compared to the vortex surface tension term in standard vortices which is of the form $\propto \bm{\nabla}\cdot\big(H_z\bm{\nabla}H_z\big)$~\cite{deGennes}, here, we have instead the appearance of the magnetic induction due to the fact that $B_z(\bm{r})\neq H_z(\bm{r})$. The second term is not present for standard magnetic-field-induced vortices and emerges only due to the magnetic island. This term corresponds to a bulk contribution and reflects the energy which is gained due to the coupling of the magnetic induction to the background spin moment of the island.

The total energy of the vortex ground state is obtained by integrating the above energy density in the interval $\rho\in[\xi_S,\infty]$. The integration of the first term in Eq.~\eqref{eq:VortexEnergy0} is given by a surface integral at $\rho=\xi_S$ and $\rho=\infty$. However, the surface contribution from $\rho=\infty$ vanishes, since the fields are zero there. We thus have:
\bea
\frac{E_{\rm vortex}}{\pi\lambda_L^2}&=&-\xi_S\left[B_z(\rho)\frac{dH_z(\rho)}{d\rho}\right]_{\rho=\xi_S}\no\\
&&+\varGamma\int_{\xi_S}^{\infty}\frac{d\rho\,\rho}{\lambda_L^2}\,I_z(\rho)B_z(\rho)\,,
\label{eq:TotalVortexEnergy0} 
\eea

\noi where we obtained the above result by accounting for the fact that all the involved functions depend only on the radial coordinate. This further allowed us to carry out the trivial integral over the angular real space coordinate.

Although the evaluation of the term in the first row of Eq.~\eqref{eq:TotalVortexEnergy0} is straightforward, it becomes more transpa\-rent by taking into account that throughout this work we assume $\xi_S\ll\lambda_L$ and $\xi_S\ll\rho_I$. Under these assumptions, when $\rho\approx\xi_S$ it is eligible to approximately replace the Bessel functions $K_0(\rho/\lambda_L)$ and $K_0(\rho/\rho_I)$ by $\ln\big(\lambda_L/\rho\big)$ and $\ln\big(\rho_I/\rho\big)$, respectively. By making these approximations, we find that:
\begin{align}
\left.-\xi_S\frac{dH_z(\rho)}{d\rho}\right|_{\rho=\xi_S}\approx\nu_\phi H_0\,.
\end{align}

On the other hand, to obtain the bulk contribution to the total vortex ground state energy, we make use of known results for the integrals of Bessel functions. See Appendix~\ref{app:AppendixA} for more details. First, we use the result:
\begin{align}
\int_{\xi_S\ll a}^\infty d\rho\,\rho\, K_0^2(\rho/a)\approx\frac{a^2}{2}\,.\label{eq:Integral1Approx}
\end{align}

\noi The second integral that becomes relevant is:
\begin{align}
\int_{\xi_S\ll a,b}^\infty d\rho\,\rho\, K_0(\rho/a)K_0(\rho/b)\approx\frac{(ab)^2}{a^2-b^2}\ln\big(a/b\big)\,,
\label{eq:Integral2Approx}
\end{align}

\noi which holds for $a\neq b$. Note that by taking the limit $b\rightarrow a$ in Eq.~\eqref{eq:Integral2Approx}, one recovers Eq.~\eqref{eq:Integral1Approx} as expected.

After taking into account the above approximate expressions, we find the final results for the dimensionless surface and area contributions:
\bea
\frac{E_{\rm vortex}^{\rm surface}}{\Phi_0H_0}&=&\frac{D_\phi}{2}\nu_\phi\left\{\nu_\phi-\nu_I\left[\frac{D_I}{D_\phi}\left(\frac{\lambda_L}{\rho_I}\right)^2
\right.\right.\no\\
&&\qquad\quad
\left.\left.-\left(1-\frac{D_I}{D_\phi}\right)\frac{1}{1-\big(\rho_I/\lambda_L\big)^2}\right]\right\},\quad
\label{eq:FinalSurfaceTotalVortexEnergy0} 
\\
\frac{E_{\rm vortex}^{\rm area}}{\Phi_0H_0}&=&
\frac{\nu_I^2/2}{1-\big(\rho_I/\lambda_L\big)^2}\left[\frac{D_\phi-D_I}{1-\big(\rho_I/\lambda_L\big)^2}-\frac{1}{2}\left(\frac{\lambda_L}{\rho_I}\right)^2\right]\no\\
&&+\frac{D_\phi}{2}\nu_\phi\nu_I\left(1-\frac{D_I}{D_\phi}\right)\frac{1}{1-\big(\rho_I/\lambda_L\big)^2}\,,
\label{eq:FinalAreaTotalVortexEnergy0} 
\eea

\noi where we introduced the stiffnesses $D_\phi=\ln\big(\lambda_L/\xi_S\big)$ and $D_I=\ln\big(\rho_I/\xi_S\big)$.

From the above contributions, the most important are the ones that involve $\nu_\phi$, since the value of $\nu_\phi$ is required to be obtained by the minimization of the energy. By ta\-king into account that we can discard terms that do not contain $\nu_\phi$, since these only yield an overall energy offset for a given $\nu_I$, we can express the relevant part that we denote $E_{\rm vortex}^{\nu_\phi}$ in the following form:
\begin{align}
\frac{E_{\rm vortex}^{\nu_\phi}}{\Phi_0H_0}=
D_\phi\frac{\big(\nu_\phi-\zeta\nu_I\big)^2}{2}\,.
\label{eq:Evortex}
\end{align}

\noi In the above, we introduced the spin-to-vorticity conversion factor:
\begin{align}
\zeta
=\frac{1}{2}\frac{D_I}{D_\phi}\left(\frac{\lambda_L}{\rho_I}\right)^2-\left(1-\frac{D_I}{D_\phi}\right)\frac{1}{1-\big(\rho_I/\lambda_L\big)^2}.
\label{eq:zeta0}
\end{align}

\noi This parameter controls the effective number of vorticity quanta $\nu_{\rm ind}=\zeta\nu_I$ which become induced by the magnetic island. In the same spirit as in the Little-Parks effect~\cite{LittleParks}, the value of $\nu_\phi$ which minimizes the energy is given by the integer which is closest to $\nu_{\rm ind}$. By relying on the structure of Eq.~\eqref{eq:zeta0}, we conclude that it is the ratio $\rho_I/\lambda_L$ rather than $D_I/D_\phi$ which primarily sets the value of the induced flux quanta $\nu_{\rm ind}$.

At this stage, it is important to clarify whether a vortex of a higher winding is stable and thus experimentally observable. For instance, Abrikosov vortices carrying vorticity values higher than unity are known to be unstable, due to the $\propto\nu_\phi^2$ term that governs their energy~\cite{deGennes}. Indeed, the latter dependence implies that a vortex carrying two units of flux will decompose into two vortices carrying a single unit of vorticity. It is straightforward to prove that such a decomposition law applies also here. Indeed, as we show in Ref.~\onlinecite{CommentOnDecomposition}, the addition of the source term $\propto\nu_I$ does not modify this property. Hence, also in the present case it is always favorable to decompose vortices of a higher winding number $\nu_\phi$ down to a number of $|\nu_\phi|$ vortices carrying a single unit of vorticity ${\rm sgn}(\nu_\phi)$.

Before proceeding with a numerical investigation of the above result, we observe that for $\rho_I\gg\lambda_L$, we obtain the approximate form:
\begin{align}
\zeta_{\rho_I\gg\lambda_L}\approx \left(1-\frac{1}{2}\frac{D_I}{D_\phi}\right)\left(\frac{\lambda_L}{\rho_I}\right)^2\,.
\label{eq:Limitrholargerlambda}
\end{align}

\noi From the above, we find that in the limit $\rho_I/\lambda_L\rightarrow\infty$ the parameter $\zeta$ tends to zero. Hence, in this regime, only a small number of vorticity quanta can be stabilized. This, however, should not be necessarily considered as a drawback. This is because for engineering vortex-MZMs, it is required to pin a superconducting vortex carrying an odd number of vorticity units. Ideal\-ly, it is desirable to establish a vortex of a single vorti\-ci\-ty unit in order to minimize the number of energetically-low-lying topologically-unprotected in-gap modes~\cite{Prada}. These are unwanted since they may be experimentally mistaken for truly topologically-protected MZMs, while they may contribute to quasiparticle poisoning that can introduce noise and suppress the fidelity of quantum manipulations.

\begin{figure}[t!]
\begin{center}
\includegraphics[width=0.95\columnwidth]{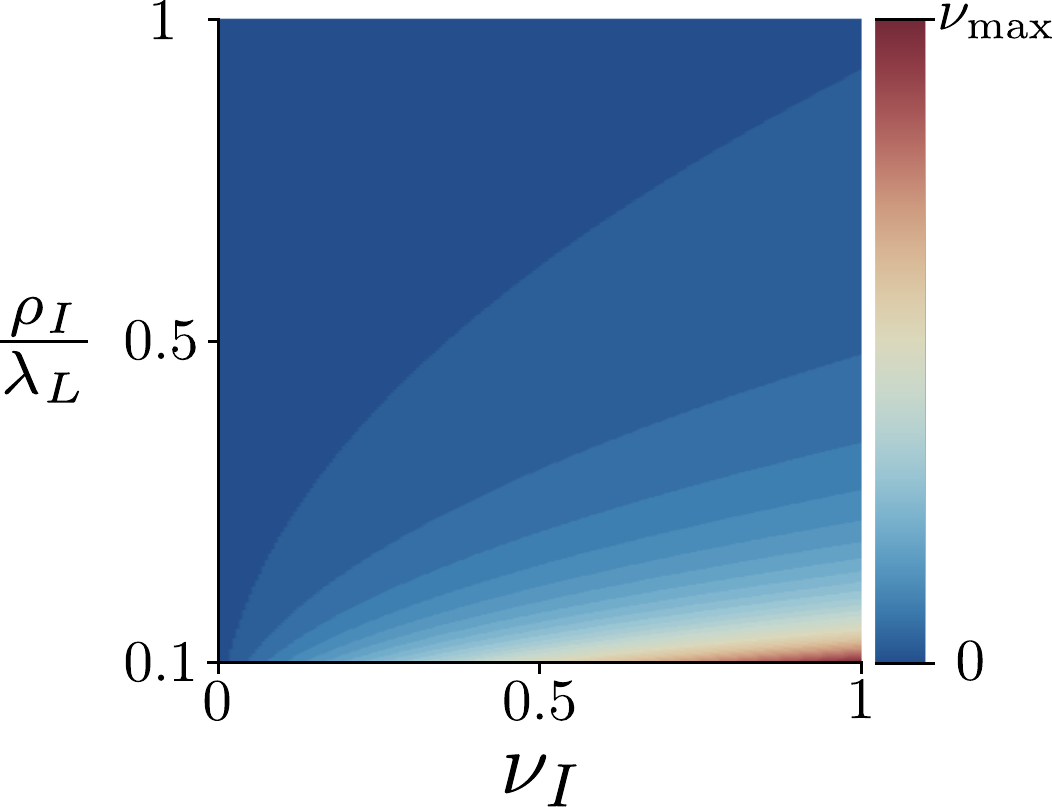}
\end{center}
\caption{The heat map shows the induced vorticity $\nu_\phi$ of a superconducting vortex stabilized in the presence of a magnetic island. Here, possible magnetic correlations are not included. The above result is obtained for $\lambda_L=1500
\,\xi_S$, in which case $\nu_{\rm max}=34$. In accordance with  our approximate analytical results in Eqs.~\eqref{eq:Limitrholargerlambda}-\eqref{eq:Limitrhosmallerlambda}, the most prominent regime for the island to induce a large number of vorticity quanta appears for $\rho_I\ll\lambda_L$, while for $\rho_I\gg\lambda_L$ the vorticity $\nu_\phi$ tends to zero.}
\label{fig:Figure2}
\end{figure} 

From the above, we conclude that, unless a small number of vorticity units is desirable, the radius of the magnetic island should not be much larger than the London penetration depth for the vortex stabilization mechanism to work efficiently. In the case where $\rho_I\rightarrow\lambda_L$, we find:
\begin{align}
\zeta_{\rho_I=\lambda_L}=\frac{1}{2}\left(1-\frac{1}{D_\phi}\right)\,.
\end{align} 

\noi We note that in the present case, additionally considering the limit $\lambda_L\rightarrow\infty$ yields that $\zeta_{\rho_I=\lambda_L}\rightarrow1/2$.

We now proceed by examining the remaining limiting scenario, i.e., the case $\lambda_L\gg\rho_I$. In this event, the spin-to-vorticity conversion factor becomes:
\bea
\zeta_{\lambda_L\gg\rho_I}\approx\frac{1}{2}\frac{D_I}{D_\phi}\left(\frac{\lambda_L}{\rho_I}\right)^2\,,
\label{eq:Limitrhosmallerlambda}
\eea

\noi which reveals that the conversion factor and the induced vorticity can reach high values in this parameter regime.

In Fig.~\ref{fig:Figure2} we present numerical results for the ari\-sing vorticity where we employ Eq.~\eqref{eq:zeta0} without any approximations. We find that our numerical results are in agreement with our approximate analytical predictions. Moreover, in Appendix~\ref{app:AppendixB}, we also calculate $\zeta$ in the case of a disk-like spatial profile $I_z(\rho)=\big(S_z/\pi\rho_I^2\big)\Theta(\rho_I-\rho)$ when considering the limits $\rho_I\gg\lambda_L$ and $\rho_I\ll\lambda_L$. In both cases, we recover the $(\lambda_L/\rho_I)^2$ dependence,  which simply reflects that $\zeta$ is proportional  to the ratio of the exchange energy $\propto S_z/\rho_I^2$ over the magnetic energy $\propto \Phi_0/\lambda_L^2$. Noteworthy, the two results map to each other for $D_I=2$.  Hence, choosing a different profile appears to solely modify the slowly-varying prefactor entering $\zeta$.

\section{Vortex Ground State in the Presence of Magnetic Correlations}\label{sec:GL}

We now proceed by incorporating the effects of magnetic correlations, and the feedback of the electronic magnetization to the vortex formation and stability. We once again adopt a phenomenological GL approach and invoke the same assumptions discussed in the previous sections. Furthermore, we consider that the superconducting co\-he\-ren\-ce length $\xi_S$ is much smaller than the correlation length $\xi_M$ gover\-ning the magnetic correlations. By virtue of this assumption, it remains eligible to continue to focus on the vortex physics outside the core.

\subsection{Ginzburg-Landau Theory}

Given the above considerations, we now add to our GL description the effects of the electronic magnetization $M_z(\bm{r})$, which becomes nonzero only after the magnetic island is added. Here, the electrons are exchange-coupled to the following li\-near combination of fields $ M_z(\bm{r})+I_z(\bm{r})-g\mu_B\hat{\bm{z}}\cdot\big[\bm{\nabla}\times\bm{a}(\bm{r})\big]/2$. In the remainder, we obtain the GL functional by redefining the magnetization field according to $M_z(\bm{r})\mapsto M_z(\bm{r})-I_z(\bm{r})$. Such a re\-de\-fi\-ni\-tion of the magnetization is allowed, since this field is determined in a self-consistent fa\-shion by minimizing the GL functional. After this shift, the electrons become exchange-coupled only to the term $M_z(\bm{r})-g\mu_B\hat{\bm{z}}\cdot\big[\bm{\nabla}\times\bm{a}(\bm{r})\big]/2$, while they are simultaneously under the additional influence of the gauge inva\-riant vector potential $\bm{a}(\bm{r})$ through the orbital coupling.

The GL functional is obtained by expanding the ener\-gy of the system up to second order with respect to the redefined magnetization field, while we also restrict to contributions which are up to second order in terms of the spatial gradients $\bm{\nabla}M_z(\bm{r})$.
By following the above steps, we obtain the intensive GL ener\-gy functional given by the expression:
\bea
\widetilde{E}_{\rm GL}(\bm{r})
&=&\frac{B_z^2(\bm{r})}{2}+\frac{\bm{a}^2(\bm{r})}{2\lambda_L^2}+\varGamma\frac{\bm{a}(\bm{r})\cdot\big[\bm{\nabla}\times\hat{\bm{z}} M_z(\bm{r})\big]}{2}\no\\
&+&\varGamma\frac{M_z(\bm{r})\hat{\bm{z}}\cdot\big[\bm{\nabla}\times\bm{a}(\bm{r})\big]}{2}-M_z(\bm{r})I_z(\bm{r})/V
\no\\
&+&\left(1/V-\chi_\perp^{\rm spin}\right)\frac{M_z^2(\bm{r})}{2}+c_M
\frac{\big[\bm{\nabla}M_z(\bm{r})\big]^2}{2}.\qquad
\label{eq:GL}
\eea

It is important to remark that two conditions hold throughout for the above GL functional. The first is $c_M\geq0$ and concerns the coefficient $c_M$ which determines the stiffness for spatial magnetic fluctuations. The se\-cond con\-di\-tion is $1/V>\chi_\perp^{\rm spin}$, and ensures that the SC is not magnetic in the absence of the magnetic island. Here, $V\geq0$ denotes the strength of a Hubbard-type  interaction which we assume to be substantial in the SC. Due to the non-negative nature of $V$, this interaction is attractive in the magnetic channel. The effects of this interaction enter through the term $M_z^2(\bm{r})/2V$ which is obtained after applying mean-field theory. As we observe from the last row in Eq.~\eqref{eq:GL}, this term provides the coupling $1/V$ between the magnetization and the spin moment field, and results from the redefinition of $M_z(\bm{r})$ discussed earlier. The functional in Eq.~\eqref{eq:GL} smoothly connects to the case where correlations become negligible by setting $V=0$. In this special case, $M_z(\bm{r})$ is solely dictated by the terms $\propto1/V$, which set it equal to $I_z(\bm{r})$.

\subsection{Vortex Solution}
\label{sec:EOM}

Along the same lines of the analysis of the vortex pinning in the absence of magnetic correlations, we begin our discussion by identifying the electric current density. In complete analogy to Eq.~\eqref{eq:Current0}, we have:
\begin{align}
\bm{J}(\bm{r})=-\varGamma\bm{\nabla}\times\hat{\bm{z}}M_z(\bm{r})-\bm{a}(\bm{r})/\lambda_L^2.
\label{eq:Current}
\end{align}

\noi Therefore, we observe that the background field $I_z(\bm{r})$ in Eq.~\eqref{eq:Current0} becomes replaced by the redefined magnetization $M_z(\bm{r})$ in Eq.~\eqref{eq:Current}. In a similar fashion, the results shown in Eqs.~\eqref{eq:Current0}-\eqref{eq:HzFull0} remain the same albeit for the substitution mentioned above. For instance, the magnetic field $H_z(\bm{r})$ still satisfies Eq.~\eqref{eq:curlHz0}, with the difference that we now have the accordingly modified definition:
\begin{align}
H_z(\bm{r})\equiv B_z(\bm{r})+\varGamma M_z(\bm{r})\,.
\end{align}

\noi By extremizing the GL functional with respect to $H_z(\bm{r})$ and $M_z(\bm{r})$, we find that in the vortex state these two fields satisfy the system of coupled EOM:
\bea
\alpha_M\left[1-\big(\xi_M\bm{\nabla}\big)^2\right]M_z(\bm{r})+\varGamma H_z(\bm{r})&=&S_M(\bm{r}),\ph
\label{eq:MzOuter}\\
-\varGamma M_z(\bm{r})+\left[1-\big(\lambda_L\bm{\nabla}\big)^2\right]H_z(\bm{r})&=&S_H(\bm{r}),\label{eq:HzOuter}
\eea

\noi where we introduced the source fields:
\bea
S_M(\bm{r})&=&\frac{1}{V}\,I_z(\bm{r})+\varGamma \nu_\phi\Phi_0\delta(\bm{r})\,,
\label{eq:S_M}\\
S_H(\bm{r})&=&\nu_\phi\Phi_0\delta(\bm{r})\,,\label{eq:S_H}
\eea

\noi and the magnetic correlation length $\xi_M=\sqrt{c_M/\alpha_M}$.  In the above, we also introduced the following quantity:
\begin{align}
\alpha_M=\frac{1}{V}-\chi_\perp^{\rm spin}-\varGamma^2\,,   
\label{eq:alphaM}
\end{align}

\noi that we also require to satisfy the constraint $\alpha_M>0$, so that the SC remains nonmagnetic even in the pre\-sen\-ce of the magnetic island. In addition, we remind the reader that the term $\nu_\phi\Phi_0\delta(\bm{r})$ drops out from Eqs.~\eqref{eq:S_M} and~\eqref{eq:S_H} in the regime outside the vortex core, which is the regime of interest in this work.

The solutions to the system of Eqs.~\eqref{eq:MzOuter} and~\eqref{eq:HzOuter} are given using the 2D Fourier transform $f(\bm{r})=\int d\bm{q}\ph e^{i\bm{q}\cdot\bm{r}}f(\bm{q})/(2\pi)^2$, and the following input functions:
\bea
M_z(\bm{q})&=&\frac{\lambda_L^2\big(q^2+1/\lambda_L^2\big)S_M(\bm{q})-\varGamma S_H(\bm{q})}{{\cal D}(q)}\,,\label{eq:MzFourier}\\
H_z(\bm{q})&=&\frac{\alpha_M\xi_M^2\big(q^2+1/\xi_M^2\big)S_H(\bm{q})+\varGamma S_M(\bm{q})}{{\cal D}(q)}\,,\ph\quad
\label{eq:HzFourier}
\eea

\noi with $\bm{q}=(q_x,q_y)$ and $q=|\bm{q}|$. The denominator reads as: 
\begin{align}
{\cal D}(q)=\alpha_M\big(\xi_M\lambda_L\big)^2\big(q^2+1/\xi_M^2\big)\big(q^2+1/\lambda_L^2\big)+\varGamma^2,\no
\end{align}

\noi and can be compactly rewritten according to the form:
\begin{align}
{\cal D}(q)\equiv\alpha_M\big(\xi_M\lambda_L\big)^2\big(q^2+1/\rho_+^2\big)\big(q^2+1/\rho_-^2\big)\,.
\end{align}

\noi In the above, we introduced two lengthscales  $\rho_\pm$, which emerge from the mixing of $M_z$ and $H_z$, and are given by the formulas:
\begin{align}
\frac{1}{\rho_\pm^2}=
\frac{1/\xi_M^2+1/\lambda_L^2}{2}\pm\sqrt{\left(\frac{1/\xi_M^2-1/\lambda_L^2}{2}\right)^2-\left(\frac{{\cal G}}{\xi_M\lambda_L}\right)^2}.
\end{align}

\noi These satisfy $\rho_-\geq\rho_+$ by definition. In the above, we also introduced the dimensionless coupling constant:
\begin{align}
{\cal G}=\frac{\varGamma}{\sqrt{\alpha_M}}\equiv\sqrt{\frac{\mu_0}{w\alpha_M}}\,{\cal X}\,,\end{align}

\noi that we can further decompose according to ${\cal G}=g_Z+g_R$, where we introduced the partial dimensionless couplings:
\begin{align}
g_{Z,R}=\frac{\gamma_{Z,R}}{\sqrt{\alpha_M}}\equiv\sqrt{\frac{\mu_0}{w\alpha_M}}\,\chi_{Z,R}\,.
\end{align}

\noi Note that our GL framework is valid as long as $\rho_\pm$ are both real, which restricts $|{\cal G}|$ to take values in the interval: 
\begin{align}
0\leq|{\cal G}|\leq|{\cal G}|_{\rm max}\quad{\rm with}\quad|{\cal G}|_{\rm max}=\frac{|\lambda_L^2-\xi_M^2|}{2\xi_M\lambda_L}\,.
\end{align}

To facilitate the inversion of the Fourier transforms in Eqs.~\eqref{eq:MzFourier} and~\eqref{eq:HzFourier}, we choose once again the spatial profile in Eq.~\eqref{eq:SpinMomentProfile}. As a result, the final outcomes for $M_z(\bm{r})$, $H_z(\bm{r})$, and $B_z(\bm{r})$ can be expressed as a sum over functions $K_0(\rho/\rho_s)$ where the set of characteristic lengthscales are identified with $\rho_s=\{\rho_I,\rho_\pm\}$. Indeed, by means of standard inverse Hankel transforms we find that the three functions of interest are given in a closed form according to the exact expression: \begin{align}
F_z(\rho)=H_0\sum_{s=I,\pm}\Big(n_IF_z^{I,s}+\nu_\phi F_z^{\phi,s}\Big)K_0\big(\rho/\rho_s\big),
\label{eq:Fformula}
\end{align}

\noi where $F_z(\rho)=\big\{{\cal M}_z(\rho),
H_z(\rho),B_z(\rho)\big\}$. In the above, we introduced the rescaled magnetization:
\begin{align}
{\cal M}_z(\rho)=\sqrt{\alpha_M}M_z(\rho),
\end{align} 

\noi which implies that $B_z(\rho)=H_z(\rho)-{\cal G}{\cal M}_z(\rho)$. To further simplify the analysis, we also define the coupling constants (with $g_I$ carrying the same dimensions as $\varGamma$):
\begin{align}
g_I=\frac{1/V}{\sqrt{\alpha_M}}\qquad{\rm and}\qquad n_I=g_I\frac{S_z}{\Phi_0}\,,
\end{align}

\noi with the dimensionless parameter $n_I$ being proportional to the number of flux quanta $\tilde{\nu}_I$ which are induced by the magnetic island in the presence of magnetic correlations.

To obtain the expressions for ${\cal M}_z(\rho)$ and $H_z(\rho)$ from Eq.~\eqref{eq:Fformula}, we need to spe\-ci\-fy the coefficients ${\cal M}_z^{I/\phi,s}$ and $H_z^{I/\phi,s}$, where $s=I,\pm$. Straightforward calculations yield ${\cal M}_z^{\phi,I}=H_z^{\phi,I}=0$, along with the exact analytical expressions for the remaining coefficients given by:
\bea
{\cal M}_z^{I,\pm}&=&\frac{1}{1+{\cal G}^2}\frac{\lambda_L^2\big(\rho_\pm^2-\lambda_L^2\big)}{\big(\rho_\pm^2-\rho_\mp^2\big)\big(\rho_\pm^2-\rho_I^2\big)},\,
\label{eq:CoefficientsFirst}\\
{\cal M}_z^{I,I}&=&\frac{1}{1+{\cal G}^2}
\frac{\lambda_L^2\big(\rho_I^2-\lambda_L^2\big)}{\big(\rho_I^2-\rho_+^2\big)\big(\rho_I^2-\rho_-^2\big)}
,\,\\
{\cal M}_z^{\phi,\pm}&=&
\frac{{\cal G}}{1+{\cal G}^2}\,\frac{\big(\lambda_L/\rho_\pm\big)^2}{\big(\rho_\mp/\lambda_L\big)^2-\big(\rho_\pm/\lambda_L\big)^2},\,\\
H_z^{I,\pm}&=&\frac{{\cal G}}{1+{\cal G}^2}\frac{\big(\lambda_L\rho_\pm\big)^2}{\big(\rho_\pm^2-\rho_\mp^2\big)\big(\rho_\pm^2-\rho_I^2\big)},\,\quad\\
H_z^{I,I}&=&\frac{{\cal G}}{1+{\cal G}^2}\frac{\big(\lambda_L\rho_I\big)^2}{\big(\rho_I^2-\rho_+^2\big)\big(\rho_I^2-\rho_-^2\big)},\,\\
H_z^{\phi,\pm}&=&\frac{1}{1+{\cal G}^2}\frac{1+{\cal G}^2-\big(\xi_M/\rho_\pm\big)^2}{\big(\rho_\pm/\lambda_L\big)^2-\big(\rho_\mp/\lambda_L\big)^2},\label{eq:CoefficientsLast}
\eea

\noi where the above expressions were obtained after making use of the relation
$(\rho_+\rho_-/\xi_M)^2=\lambda_L^2/(1+{\cal G}^2)$. 

As it is customary, in order to study the possible emergence of a nonzero phase winding $\nu_\phi$, it is required to first obtain the ener\-gy for the vortex solution found above. This is presented in the upcoming section. 

\subsection{Vortex Stability and Phase Winding} 
\label{sec:VortexStability}

An expression for the energy at the extremum obtained by solving the EOM, is found by plugging Eqs.~\eqref{eq:MzOuter} and~\eqref{eq:HzOuter} into the GL functional of Eq.~\eqref{eq:GL}. To facilitate this procedure and render it more transparent, we express Eqs.~\eqref{eq:GL},~\eqref{eq:MzOuter} and~\eqref{eq:HzOuter} in terms of the magnetic field $H_z(\bm{r})$ and the rescaled magnetization ${\cal M}_z(\bm{r})$. After carrying out the required substitutions, we find the following form for the GL functional:
\bea
&&\widetilde{E}_{\rm GL}(\bm{r})=\frac{{\cal M}_z^2(\bm{r})}{2}+\frac{\big[\xi_M\bm{\nabla}{\cal M}_z(\bm{r})\big]^2}{2}-g_I{\cal M}_z(\bm{r})I_z(\bm{r})\no\\
&&+\frac{H_z^2(\bm{r})}{2}+\frac{\big[\lambda_L\bm{\nabla}H_z(\bm{r})\big]^2}{2}-\frac{\lambda_L^2}{2}\bm{\nabla}\cdot\big[{\cal G}{\cal M}_z(\bm{r})\bm{\nabla}H_z(\bm{r})\big],\no\\
\label{eq:GLnew}
\eea

\noi along with the following expressions for the two EOM in the region $\rho>\xi_S$:
\bea
\left[1-\big(\lambda_L\bm{\nabla}\big)^2\right]H_z(\bm{r})&=&+{\cal G}{\cal M}_z(\bm{r})\,,\\
\label{eq:HzFull}
\left[1-\big(\xi_M\bm{\nabla}\big)^2\right]{\cal M}_z(\bm{r})&=&-{\cal G}H_z(\bm{r})+g_II_z(\bm{r}).\label{eq:MzFull}
\eea

\begin{figure*}[t!]
\begin{center}
\includegraphics[width=1\textwidth]{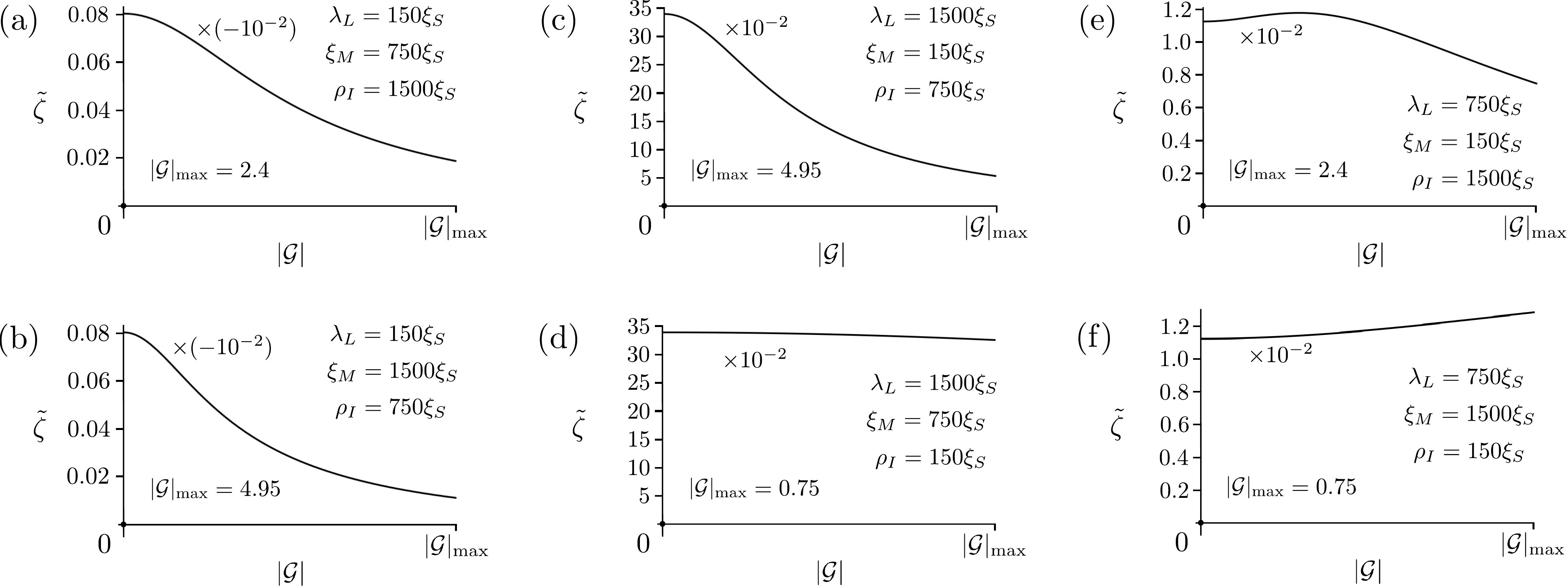}
\end{center}
\caption{Spin-to-flux conversion factor $\tilde{\zeta}$ as a function of the absolute value of the coupling strength $|{\cal G}|\in\big[0,|{\cal G}_{\rm max}|\big]$. Here, we focus on the six possible hierarchies for the lengthscales $\{\lambda_L,\xi_M,\rho_I\}$ which are obtained when these three quantities are substantially different. We find that the conversion efficiency becomes the highest (lowest) for $\lambda_L\gg\xi_M,\rho_I$ ($\lambda_L\ll\xi_M,\rho_I$), as this is reflected in panels (c)-(d) ((a)-(b)). Intermediate values of $\tilde{\zeta}$ are correspondingly obtained when $\lambda_L$ is placed in-between $\xi_M$ and $\rho_I$. See panels (e)-(f). We observe that the enhancement of the mixing between the magnetic and magnetization fields does not boost the conversion efficiency. In fact, it greatly suppresses it in all cases except for those in which $\rho_I\ll\lambda_L,\xi_M$. Approximate analytical expressions for $\tilde{\zeta}$ for weak ($|{\cal G}|=0$) and strong ($|{\cal G}|=|{\cal G}|_{\rm max}$) couplings are given in Appendix~\ref{app:AppendixStrong2Weak}. The value of $\tilde{\zeta}$ for each hierarchy is obtained after multiplying the values in the vertical axis by the factor shown in each graph.}
\label{fig:Figure3}
\end{figure*}

By employing the above two relations, we obtain the energy contribution of the outer part of the vortex core when con\-si\-de\-ring the vortex ground state extremum:
\bea
\widetilde{E}_{\rm vortex}(\bm{r})&=&
\frac{\lambda_L^2}{2}\bm{\nabla}\cdot\big[B_z(\bm{r})\bm{\nabla}H_z(\bm{r})\big]-\frac{1}{2}g_II_z(\bm{r}){\cal M}_z(\bm{r})\no\\
&+&
\frac{\xi_M^2}{2}\bm{\nabla}\cdot\big[{\cal M}_z(\bm{r})\bm{\nabla}{\cal M}_z(\bm{r})\big].
\label{eq:VortexEnergy}
\eea

\noi Noteworthy, the first two terms have a one-to-one correspondence to the first two terms appearing in Eq.~\eqref{eq:VortexEnergy0}. Here, one finds an additional ener\-gy tension term for the vortex formation, i.e., the last contribution, which stems from the build up of the electronic magnetization.

By carrying out the spatial integration and putting together the various terms, we obtain:
\begin{align}
\frac{\widetilde{E}_{\rm vortex}^{\nu_\phi}}{\Phi_0H_0}=\widetilde{D}_\phi\frac{\big(\nu_\phi-\tilde{\zeta}\tilde{\nu}_I\big)^2}{2}\,.
\label{eq:Etildevortex}
\end{align}

\noi Notably, in the presence of magnetic correlations the vorticity stiffness becomes:
\begin{align}
\widetilde{D}_\phi=\sum_{s,s'}^\pm D_s\left[B_z^{\phi,s}H_z^{\phi,s'}+\left(\frac{\xi_M}{\lambda_L}\right)^2{\cal M}_z^{\phi,s}{\cal M}_z^{\phi,s'}\right],
\label{eq:StiffnessCor}
\end{align}

\noi which is expressed in terms of the partial stiffnesses $ D_\pm=\ln\big(\rho_\pm/\xi_S\big)$. In addition, in Eq.~\eqref{eq:Etildevortex} we introduced the modified spin-to-vorticity conversion factor:
\begin{align}
\tilde{\zeta}=\sum_s^\pm\Big(D_s/\widetilde{D}_\phi-D_I/\widetilde{D}_\phi\Big)\frac{1}{1-\big(\rho_I/\rho_s\big)^2}\,\frac{{\cal M}_z^{\phi,s}}{2{\cal G}}\qquad\quad\phd\no\\
-\sum_{s,s'}^{\pm,I}\frac{D_s}{\widetilde{D}_\phi}\Bigg[\frac{B_z^{I,s}H_z^{\phi,s'}}{2{\cal G}}+\left(\frac{\xi_M}{\lambda_L}\right)^2\frac{{\cal M}_z^{I,s}{\cal M}_z^{\phi,s'}}{2{\cal G}}+I\leftrightarrow\phi\Bigg].\end{align}

With the above expressions at hand, we are now in a position to define the quantity $\tilde{\nu}_{\rm ind}=\tilde{\zeta}\tilde{\nu}_I$ which is the analog of $\nu_{\rm ind}$ and, thus, sets the number of vorti\-ci\-ty quanta which become stabilized for $\tilde{\nu}_I$ units of flux induced by the island. In the present case we find that:
\begin{align}
\tilde{\nu}_I={\cal G}n_I=\frac{\nu_I}{V\alpha_M}=\frac{\nu_I}{1-V\chi_\perp^{\rm spin}-V\varGamma^2}\equiv\frac{\big(1+{\cal G}^2\big)\nu_I}{1-V\chi_\perp^{\rm spin}}\,.
\label{eq:FluxQuanta}
\end{align}

\noi To obtain the last expression, we made use of the fact that after replacing $\Gamma$ in terms of ${\cal G}$, Eq.~\eqref{eq:alphaM} takes the form $V\alpha_M=1-V\chi_\perp^{\rm spin}-{\cal G}^2\alpha_M\Rightarrow \big(1+{\cal G}^2\big)V\alpha_M=1-V\chi_\perp^{\rm spin}$. Rewriting $\tilde{\nu}_I$ as in Eq.~\eqref{eq:FluxQuanta} is convenient when examining general properties of the vortex formation and treating ${\cal G}$ as an independent parameter.

From the result in Eq.~\eqref{eq:FluxQuanta}, we conclude that the first consequence of the presence of the magnetic correlations is a renormalization of the vorticity quanta induced by the magnetic island. Notably, as the system gets closer to the magnetic phase transition, $\tilde{\nu}_I$ can become signi\-fi\-cant\-ly larger than $\nu_I$. However, aside from modifying the number of induced vorticity quanta, the magnetic correlations also drastically affect the structure of the remaining two key physical quantities which dictate the vortex formation, i.e., the vorticity stiffness $\widetilde{D}_\phi$ and the spin-to-vorticity conversion factor $\tilde{\zeta}$.

\subsection{Correlations and Spin-to-Flux Conversion}\label{sec:Spin2Flux}

In Fig.~\ref{fig:Figure3} we depict the results for the conversion factor $\tilde{\zeta}$ in the six hierarchies that become possible for $\{\lambda_L,\xi_M,\rho_I\}$, when these are all unequal and take substantially different values.  Panels (a)-(b) in Fig.~\ref{fig:Figure3} show the evaluated $\tilde{\zeta}$ when the London penetration depth is much smaller than the two other lengthscales. The outcomes in these two situations do not differ much, with the conversion factor being rather small in both, of the order $10^{-3}-10^{-4}$. Notably, the maximum conversion is obtained when the electromagnetic sector and the magnetic fluctuations become fully decoupled, i.e., for $|{\cal G}|=0$. In contrast, the inverse hierarchy with $\lambda_L\gg\rho_I,\xi_M$ provides larger values for $\tilde{\zeta}$, with a maximum that is approximately equal to $\nicefrac{1}{3}$. Once again, the conversion efficiency becomes reduced upon increasing $|{\cal G}|$, as one can immediately infer from Figs.~\ref{fig:Figure3}(c)-(d). Notably, this reduction is extremely weak for the hierarchy $\rho_I\ll\xi_M\ll\lambda_L$ in Fig.~\ref{fig:Figure3}(d), in which case $\tilde{\zeta}$ remains practically constant in the entire window of accessible $|{\cal G}|$ values. Lastly, as shown in Figs.~\ref{fig:Figure3}(e)-(f), when $\lambda_L$ is positioned in-between $\xi_M$ and $\rho_I$, the resulting conversion factor also ends up taking values in-between the ones found in Figs.~\ref{fig:Figure3}(a)-(b) and Figs.~\ref{fig:Figure3}(c)-(d), respectively. Therefore, we conclude that it is mainly the position of $\lambda_L$ in the hierarchy of the three relevant lengthscales that decides the magnitude of $\tilde{\zeta}$. Moreover, we find that a stronger coupling between the magnetic and magnetization fields generally disfavors the magnetic-island-pinning of superconducting vortices, with the exception being scenarios where $\rho_I\ll\lambda_L,\xi_M$ holds, since in these cases the conversion efficiency becomes essentially independent of the precise value of $|{\cal G}|$.

From the above results it becomes apparent that the weak-coupling regime is the most relevant for pinning vortices, since almost in all cases this enables to achieve the maximum spin-to-flux conversion efficiency. Hence, it is meaningful to acquire a better understanding of the limit $|{\cal G}|\ll|{\cal G}|_{\rm max}$, in which case the characteristic lengthscales $\rho_\pm$ of the mixed magnetic and magnetization fields are practically mapped to $\lambda_L$ and $\xi_M$, accor\-ding to $\rho_+\approx\xi_M$ ($\rho_+\approx\lambda_L$) and $\rho_-\approx\lambda_L$ ($\rho_-\approx\xi_M$) for $\lambda_L\gg\xi_M$ ($\xi_M\gg\lambda_L$). These simplifications allow us to obtain approximate analytical results for the conversion factor, which we discuss in Appendix~\ref{app:AppendixStrong2Weak}. Overall, we find that in all possible six hierar\-chies $\tilde{\zeta}\propto\big(\lambda_L/{\rm max}\{\rho_I,\xi_M\}\big)^2$, with a prefactor that involves the stiffnesses $D_{\phi,M,I}$ and varies depending on the case. Here, we set $D_M=\ln\big(\xi_M/\xi_S\big)$. These results are analogous to the ones found previously in Eqs.~\eqref{eq:Limitrholargerlambda} and~\eqref{eq:Limitrhosmallerlambda}, in the absence of magnetic correlations. The key difference when correlations are included is that the effective radius of the magnetic island is now set by the largest lengthscale arising from $\rho_I$ and $\xi_M$.

We conclude this section by also briefly discussing the antipodal limit, i.e., the one in which the coupling between magnetization and magnetic fields is the strongest possible. Thus, here we have the condition $|{\cal G}|=|{\cal G}|_{\rm max}$, which leads to $\rho_\pm\rightarrow\bar{\rho}$ with the common lengthscale $
1/\bar{\rho}=\sqrt{\big(1/\xi_M^2+1/\lambda_L^2\big)/2}$. The arising equality of $\rho_\pm$ leads also here to a number of simplifications. In Appendix~\ref{app:AppendixStrong2Weak} we also provide approximate expressions obtained for the conversion factor in this strong coupling limit. Our main finding related to this case is that the strong mixing of the electromagnetic sector with the magnetic fluctuations leads to radically different behaviors than the ones entailed by their non-correlated counterparts in Eqs.~\eqref{eq:Limitrholargerlambda} and~\eqref{eq:Limitrhosmallerlambda}. This is in full agreement with the overall trends observed in Fig.~\ref{fig:Figure3} and the ge\-ne\-ral\-ly dif\-fe\-rent outcomes for the spin-to-flux conversion factor in the two extreme limits $|{\cal G}|=0$ and $|{\cal G}|=|{\cal G}|_{\rm max}$.

\subsection{ Effective Picture in the Weak-Coupling Regime}\label{sec:effectivePicture}

Based on the results obtained in the previous paragraphs, we conclude that the weak coupling limit ${\cal G}=0$ is generally the most prominent for enabling a large spin-to-flux conversion. Our analysis in the upcoming sections also proves that the limit ${\cal G}\approx0$ is also the experimentally most relevant regime for pinning superconducting vortices via our mechanism in the two types of concrete Rashba systems of interest in this work. These conclusions motivate us to examine in more detail the vortex solution and the spatial profiles for the va\-rious fields in this limit. Our starting point is Eq.~\eqref{eq:Fformula} along with the results in Eqs.~\eqref{eq:CoefficientsFirst}-\eqref{eq:CoefficientsLast}. By assuming $G\approx0$ and $\lambda_L\gg\rho_I,\xi_M$, we find that aside from the known results ${\cal M}_z^{\phi,I}=H_z^{\phi,I}=0$, also the coefficients ${\cal M}_z^{I,-}$ and $H_z^{\phi,+}$ become approximately zero in this limiting case. In contrast, the re\-mai\-ning coefficients are nonzero and read as $
{\cal M}_z^{I,+}\approx\lambda_L^2/(\xi_M^2-\rho_I^2)$, ${\cal M}_z^{I,I}\approx \lambda_L^2/(\rho_I^2-\xi_M^2)$,
${\cal M}_z^{\phi,+}={\cal G}(\lambda_L/{\xi_M})^2$, $
{\cal M}_z^{\phi,-}\approx-{\cal G}$, $H_z^{I,+}\approx{\cal G}\xi_M^2/(\rho_I^2-\xi_M^2)$, $H_z^{I,I}\approx{\cal G}\rho_I^2/(\xi_M^2-\rho_I^2)$, $
H_z^{I,-}\approx{\cal G}$, and $H_z^{\phi,-}\approx1$.

We now examine the meaning of the above results when considering the original unit system and definition for $M_z(\bm{r})$. In particular, we focus on the expressions for the ``dressed'' magnetic-island field and the magnetic induction. By retaining the lowest-order terms in the coupling ${\cal G}$, we find the following results:
\bea
&&I_z(\rho)+M_z(\rho)\approx\frac{\tilde{S}_z}{2\pi}\frac{K_0(\rho/\rho_I)-K_0(\rho/\xi_M)}{\rho_I^2-\xi_M^2}\,,\\
&&\frac{B_z(\rho)}{{\cal X}\mu_0/w}\approx\big(1+\tilde{\zeta}\big)\frac{\tilde{S}_z}{2\pi\lambda_L^2}K_0(\rho/\lambda_L)
-\left[I_z(\rho)+M_z(\rho)\right].\no\\\label{eq:BforSmallG}
\eea

\noi To obtain the above expressions we considered $I_z(\rho)+M_z(\rho)$ at zeroth order in ${\cal G}$, since any corrections arise at quadratic order. For the magnetic induction we instead kept terms at first order. There exist no zeroth order contributions to the magnetic induction, since this is solely sourced by the spin moment field of the magnetic island.

In order to derive Eq.~\eqref{eq:BforSmallG}, we assumed for convenience that $\tilde{\nu}_{\rm ind}=\tilde{\zeta}\tilde{\nu}_I\in\mathbb{Z}$  with no loss of generality, and subsequently set $\nu_\phi=\tilde{\nu}_{\rm ind}$. This allowed us to express both $n_I$ and $\nu_\phi$ in terms of the modified spin moment of the island due to the presence of the magnetic correlations, i.e., in terms of $\tilde{S}_z=S_z/(V\alpha_M)$. As a result, we find that the spin moment becomes effectively renormalized by the factor $1/(V\alpha_M)$ which corresponds to the usual factor emer\-ging in the so-called random phase approxi\-ma\-tion (RPA), see for instance Ref.~\onlinecite{RPA}. This is expected, since the RPA is equivalent to the mean-field decoupling theory that has been implicitly employed throughout this work. For small ${\cal G}$, the RPA factor takes the form $1/(V\alpha_M)\approx1/(1-V\chi_\perp^{\rm spin})$ and diverges at the magnetic phase transition occurring for the critical value $V_c=1/\chi_\perp^{\rm spin}$. Hence, irrespectively of the influence that the magnetic correlations may have via introducing the lengthscale $\xi_M$, a major consequence of the presence of correlations is to ``dress'' and effectively enhance the spin moment of the magnetic island, which is an effect that generally facilitates the vortex pinning for all systems.

Having identified the above fields, we are now in a position to obtain the effective exchange field $\tilde{I}_z(\rho)=I_z(\rho)+M_z(\rho)-(g\mu_B/2)B_z(\rho)$ that the electrons of the SC are exposed to in the presence of the magnetic island. Note that, here, we restrict to $\rho>0$. If we further assume that $\xi_M$ and $\rho_I$ differ substantially, we obtain:
\bea
\tilde{I}_z(\rho)&\approx&\left(1+\frac{g\mu_B{\cal X}\mu_0}{2w}\right)\frac{\tilde{S}_z}{2\pi\tilde{\rho}_I^2}K_0\big(\rho/\tilde{\rho}_I\big)\no\\
&-&\big(1+\tilde{\zeta}\big)\frac{g\mu_B{\cal X}\mu_0}{2w}\frac{\tilde{S}_z}{2\pi\lambda_L^2}K_0(\rho/\lambda_L)\,,\label{eq:IzEffective}
\eea

\noi where we introduced the effective radius of the magnetic island $\tilde{\rho}_I={\rm max}\{\rho_I,\xi_M\}$. Note that the above approximate expression holds well for $\rho\gg{\rm min}\{\rho_I,\xi_M\}$. Moreover, $\tilde{\zeta}$ can be also replaced by a suitable approximate form that we have obtained in Appendix~\ref{app:AppendixStrong2Weak}.

As it becomes more transparent in the upcoming sections, obtaining a detailed expression for the effective spin moment field $\tilde{I}_z(\rho)$ is useful for inferring the topological properties of the Rashba SC. Therefore, it is helpful to better understand it. For this purpose, we estimate the dimensionless constant $g\mu_B{\cal X}\mu_0/(2w)$. After considering $g=2$, $w=5\,{\rm nm}$, and evaluating ${\cal X}$ using the results of Ref.~\onlinecite{Anatomy}, we find that for the superconducting TI surface states this constant is approximately $ 9\times10^{-7}$, while for the case of a superconduc\-ting Rashba metal this quantity becomes roughly three times smaller. Hence, the contributions proportional to $g\mu_B{\cal X}\mu_0/(2w)$ are negligible for the cases of interest, and the effective spin moment field of the magnetic island takes the form:\begin{align}
\tilde{I}_z(\rho)\approx\tilde{I}_zK_0\big(\rho/\tilde{\rho}_I\big),\label{eq:EffectiveIz}
\end{align}

\noi with the exchange energy:
\begin{align}
\tilde{I}_z=\frac{\tilde{S}_z}{2\pi\tilde{\rho}_I^2}.
\end{align}

\noi We remark that the contribution in the second line of Eq.~\eqref{eq:IzEffective} which is proportional to $\tilde{\zeta}$ can be safely neglected because  $\tilde{\zeta}$ is found to be proportional to $(\lambda_L/\tilde{\rho}_I)^2$ with a numerical coefficient which is of the order of one. See Appendix~\ref{app:AppendixStrong2Weak} for further details.

\section{Application to concrete systems}\label{sec:RepresentativeModels}

We now employ our analysis to infer the possible emergence of nonstandard vortices in two concrete Rashba SCs. First we obtain the values for $\xi_M$ and $\lambda_L$ dictating each Rashba system, along with the couplings $g_{Z,R}$, and ${\cal G}$. For this purpose, we consider the microscopic model in Eq.~\eqref{eq:Hamiltonian}. We provide extensive details regarding this procedure in Appendix~\ref{app:AppendixGLcoeff}. We subsequently derive a criterion for the vortex formation by means of our mechanism. By employing this criterion, we complete our analysis by providing a conclusive answer on whether our proposal is applicable or not to these two classes of materials.

\subsection{Topological Surface States}\label{sec:TIstates}

We now consider realistic values for the GL parameters obtained in Appendix~\ref{app:AppendixGLcoeff}. Our choice for these values is motivated by observations in FeTeSe compounds, since these systems have been already expe\-ri\-men\-tal\-ly claimed to harbor superconducting vortices induced by magnetic impurities~\cite{QAHVexp}. In the following, we assume that the to\-po\-lo\-gi\-cal surface states leak inside the bulk FeTeSe system within a width of $w=5\,{\rm nm}$. In addition, we consider a gyromagnetic factor $g=2$, a Fermi energy $E_F=4.5\,{\rm meV}$, a pai\-ring gap $\Delta=1.5\,{\rm meV}$, and a cutoff energy $\Lambda=20\,{\rm meV}$. See for instance Refs.~\onlinecite{KunJiang,KongDing}. Furthermore, we choose $\hbar\upsilon_R=0.2\,{\rm eV\cdot\AA}$ for the strength of the effective Rashba SOC, which in the present context results from the spin-momentum locking of the surface states. The superconducting coherence length is calculated using the expression $\xi_S=\hbar\upsilon_F/(\pi\Delta)$~\cite{deGennes}. For the above parameter va\-lues we find $\xi_S\simeq3.7\,{\rm nm}$, which is comparable to the experimentally observed value of about two nanometers~\cite{Lei}. On the other hand, we obtain the London penetration depth $\lambda_L\simeq4.9\,{\rm \mu m}$, which is of the same order of magnitude as the observed value of $1.5\,{\rm \mu m}$~\cite{Toulemonde}. Hence, our theoretical modeling is  consistent with the  experimental observations in FeTeSe. This implies that, within our description, the superconducting TI surface indeed behaves as a type-II SC which generally allows for the stabilization of superconducting vortices.

Having ensured that the desired condition $\lambda_L\gg\xi_S$ is satisfied, we examine the behavior of $\xi_M$ and ${\cal G}$ upon varying the interaction strength $V$. For this purpose, it is more convenient and transparent to re-express $V$ in terms of the dimensionless parameter $\eta$, according to:
\begin{align}
V\equiv\frac{\eta}{\chi_\perp^{\rm spin}+\Gamma^2}\,.
\end{align}

\noi From the above, we find that the interaction is given in units of $1/(\chi_\perp^{\rm spin}+\Gamma^2)\simeq175\,{\rm meV\cdot nm^2}$, while the Fermi wavelength is found to be $\lambda_F=2\pi/k_F\approx28\,{\rm nm}$. The introduction of $\eta$ allows us to write $V\alpha_M=1-\eta$. This reflects that a magnetic phase transition occurs for the critical value $\eta_c=1$, at which the magnetic correlation length $\xi_M$ diverges. The required critical interaction value $V_c$ for reaching the critical point is relatively small, since we find that $1/\big[\lambda_F^2(\chi_\perp^{\rm spin}+\Gamma^2)\big]\approx223\,{\rm \mu eV}$, which further corroborates the strong tendency of the FeTeSe compounds to exhibit a magnetic instability~\cite{FeTeSeMagnetic}. Since throughout this work we do not consider such a possibility, we restrict to the regime $\eta\in[0,1)$, with $\eta=0$ corresponding to the case where correlations are absent.

\begin{figure}[t!]
\begin{center}
\includegraphics[width=\columnwidth]{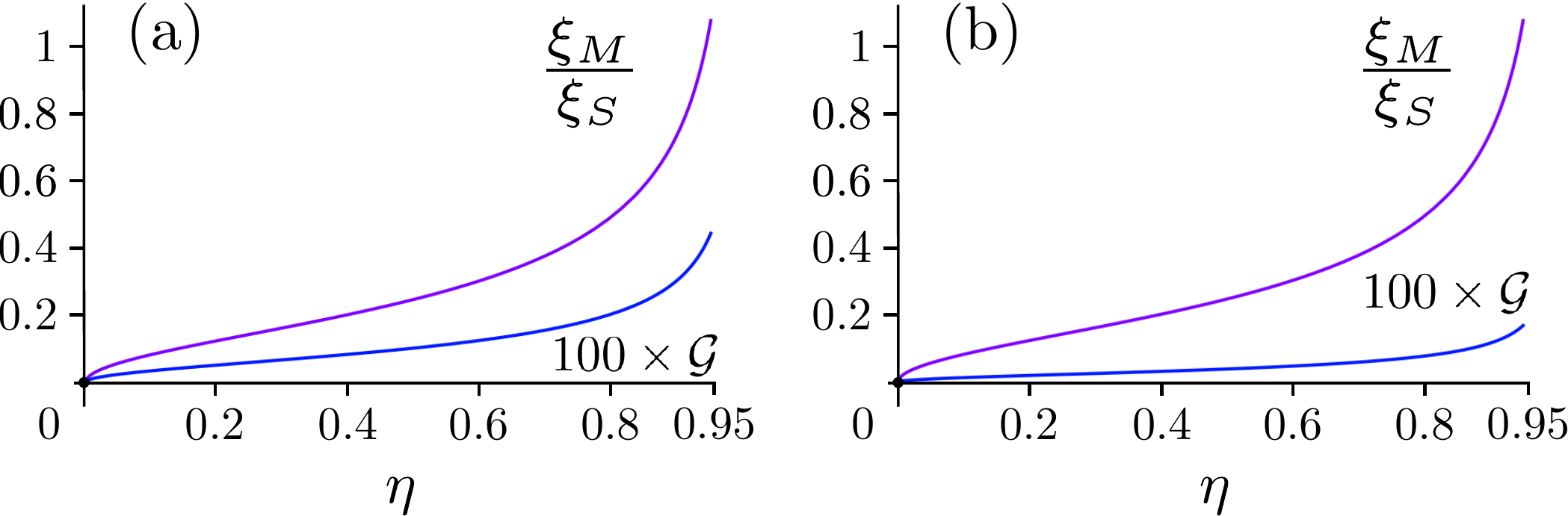}
\end{center}
\caption{Results for the parameters $\xi_M$ and ${\cal G}$ in the case of (a) disorder-free TI surface states and (b) a disordered Rashba metal. We find very similar results in both situations. The magnetic correlation length increases in terms of the pa\-ra\-me\-ter $\eta\in[0,0.95)$, which controls the strength of the magnetic interaction. A magnetic instability appears for $\eta_c=1$, which is outside the regime of interest in the present work. We find that the correlations are required to be substantial so that $\xi_M\gg\xi_S$. We also find that the coupling ${\cal G}$ increases upon increasing $\eta$ but generally remains small in a wide range of values, unless one tunes the system extremely close to the magnetic instability. Here we multiplied ${\cal G}$ by a factor of 100, in order to conveniently plot the two quantities together.}
\label{fig:Figure4}
\end{figure}

In Fig.~\ref{fig:Figure4}(a) we show results for $\xi_M$ and ${\cal G}$ when varying $\eta$ in the interval $[0,0.95)$. We observe that $\xi_M$ becomes comparable to $\xi_S$ only when the correlations become substantial. Additional calculations which are not included in Fig.~\ref{fig:Figure4}, provide that $\xi_M/\xi_S\simeq7.8$ for $\eta=0.999$. Moreover, the strength of the mixing between magnetic and magnetization fields ${\cal G}$ takes small values in the entire range, i.e., of the order of $10^{-2}-10^{-3}$, while we find that ${\cal G}_{\rm max}=906.6$ (${\cal G}_{\rm max}=86.1$) for $\eta=0.9$ ($\eta=0.999$). Therefore, the weak-coupling limit with ${\cal G}\approx0$ applies, and one can utilize the results of the previous section to infer the conditions for pinning superconducting vortices. As we earlier pointed out, the weak-coupling limit is generally favorable for induces vortices, while the spin-to-vorticity factor will be further determined by which lengthscale hier\-archy of $\{\lambda_L,\xi_M,\rho_I\}$ becomes relevant.

\subsection{Rashba Metal}\label{sec:Rashba}

We now proceed and consider concrete parameter va\-lues for the GL coefficients in the case of a Rashba metal. Here, we choose values with an eye to applying our theory to systems similar to Pb deposited on top of Si(111) surfaces that was recently experimentally stu\-died in Refs.~\onlinecite{Cren,Menard}. In  particular, we set $w=5\,{\rm nm}$, $E_F=750\,{\rm meV}$, $\Delta=1\,{\rm meV}$, and $\hbar\upsilon_R=0.2\,{\rm eV\cdot\AA}$. In the absence of the Rashba SOC, these result in a Fermi wavelength which is equal to $\lambda_F\simeq1.4\,{\rm nm}$, a Fermi ve\-lo\-ci\-ty $\upsilon_F\simeq5.1\times10^5\,{\rm m/s}$, while $\upsilon_R\simeq0.3\times10^5\,{\rm m/s}$.

Using the above, we find that the superconducting coherence length is $\xi_S\simeq108\,{\rm nm}$ and the London penetration depth becomes $\lambda_L\simeq213\,{\rm nm}$. We thus obtain the GL-number $\kappa\equiv\lambda_L/\xi_S\approx2$, which implies that the system is expected to behave as a type-II SC. However, since $\kappa$ is close to the critical GL-number $\kappa_c=1/\sqrt{2}$ se\-pa\-ra\-ting type-I from type-II SCs, we would naively not expect from such a system to be a prominent candidate for experimentally realizing our proposal. This would naturally hold for all elemental SCs which are typically dictated by a very large Fermi energy, a property that in most cases renders them type-I SCs. However, ta\-king into account the possible presence of disorder offers a loophole that allows a number of type-I elemental SCs to convert into type-II and host superconduc\-ting vortices~\cite{deGennes}. This in fact is the case for Pb-Si(111) which is substantially disordered, with a mean-free path $\ell\sim4\,{\rm nm}$~\cite{Cren,Menard}. In the presence of disorder, both coherence length and London penetration depth become affected. For a strong disorder leading to a mean-free path $\ell\ll\xi_{\rm S}$, the coherence length and penetration depth are modified according to $\xi_{S}\mapsto\xi_{S}\sqrt{\ell/\xi_{S}}$ and $\lambda_L\mapsto\lambda_L\sqrt{\xi_{S}/\ell}$, so that the GL-number changes as $\kappa\simeq\lambda_L/\xi_{S}\mapsto \kappa(\xi_{S}/\ell)$, see Ref.~\onlinecite{deGennes}.

In the remainder, we incorporate the effects of disorder by introducing a mean-free path $\ell=4\,{\rm nm}$. This, in turn, leads to the modified quantities $\xi_S\simeq21\,{\rm nm}$ and $\lambda_L\simeq1.1\,{\rm \mu m}$. Quite remarkably, disorder modifies these two variables in such a manner, so that the present situation becomes akin to the one examined in the case of the TI surface states, which features a high GL-number without requiring the presence of disorder. Before proceeding, we remark that for the following analysis of $\xi_M$ and ${\cal G}$ we do not consider any modifications due to disorder other than those discussed for $\lambda_L$ and $\xi_S$. We follow this approach because our theoretical model contains a sufficient number of free parameters, $V$, $S_z$, and $\rho_I$. Therefore, any additional effects of disorder can be incorporated in re-definitions of the above free variables.

Given the above assumptions, in Fig.~\ref{fig:Figure4}(b) we present the results for $\xi_M/\xi_S$ and ${\cal G}$ for a superconducting Rashba metal. The emerging picture is quite similar to the TI case, mainly due to the arising similarity between $\xi_S$ and $\lambda_L$ in the two cases. The only minor difference is that ${\cal G}$ is even smaller here compared to the TI case. For $\eta=0.999$ we find $\xi_M=158\,{\rm nm}$, $\xi_M/\xi_S\simeq7.6$,  $\alpha_M\simeq2.6\times10^{34}$, ${\cal G}\simeq1.2\times10^{-2}$, and ${\cal G}_{\rm max}\simeq3.4$. Here, we find that ${\cal G}=g_Z/2$ since $g_R=-g_Z/2$. Therefore, the couplings stemming from the Zeeman and Rashba magnetoelectricity are of the same order, but opposite. This is in stark contrast to what we obtain for the TI case, in which ${\cal G}\approx g_Z$, since $g_Z/g_R\approx6$ for a broad range of values for $\eta$. Note, however, that for both the TI and Rashba metal we have $\chi_\perp^{\rm spin}\gg\Gamma^2$, which is consistent with the fact that ${\cal G}\approx0$.

The unit of the interaction strength in the present case is given as $1/(\chi_\perp^{\rm spin}+\Gamma^2)\simeq240\,{\rm meV\cdot nm^2}$, while $1/\big[\lambda_F^2(\chi_\perp^{\rm spin}+\Gamma^2)\big]\approx119\,{\rm meV}$. Notably, the critical interaction for reaching the magnetic critical point is about 500 times larger than the one found to be required in the TI case. This is a direct consequence of the substantially larger Fermi energy for the Rashba system, which indicates that correlations can be less efficiently exploited for tuning the pinning of supercon\-duc\-ting vortices in the present class of systems. Nonetheless, this required interaction scale is yet not that large, which implies that the strong correlations regime is generally accessible also here. Indeed, a large Hubbard interaction of the order of $1\,{\rm eV}$ has been predicted for Pb on substrates~\cite{PbHubbard}, which hints that correlations can be relevant in these systems.

\subsection{ Criterion for Zero-Field Vortex Formation}\label{sec:SimpleCriterion}

We conclude this section by identifying the condition that needs to be satisfied in order to induce a superconducting vortex with a single unit of vorticity. For this to take place, $|\tilde{\nu}_{\rm ind}|$ needs to reach the value $\nicefrac{1}{2}$ and be smaller than $\nicefrac{3}{2}$. This implies that the critical condition for a single-unit vortex to be stabilized is $\tilde{\zeta}\tilde{\nu}_I=\nicefrac{1}{2}$, which equivalently leads to the following threshold value:
\begin{align}
\tilde{I}_z=\frac{\ln(\lambda_L/\xi_S)}{\ln\big({\rm max}\{\rho_I,\xi_M\}/{\rm min}\{\rho_I,\xi_M\}\big)}\frac{\Phi_0}{2\pi\lambda_L^2\mu_0{\cal X}/w}\,,\label{eq:VortexCriterion}
\end{align}

\noi for the exchange energy $\tilde{I}_z$ felt by the electrons due to the magnetic island. We remark that the above is expressed and derived in the original SI unit system, by employing the relations $\tilde{\nu}_I=\mu_0\tilde{S}_z{\cal X}/w\Phi_0$, $\tilde{I}_z=\tilde{S}_z/(2\pi\tilde{\rho}_I^2)$, and the approximate result for $\tilde{\zeta}$ described in
Eq.~\eqref{eq:ZetaApprox} when assuming the here-relevant weak coupling limit. We note that the first fraction on the r.h.s. of Eq.~\eqref{eq:VortexCriterion} depends only on the ratios $\rho_I/\xi_M$ and $\lambda_L/\xi_S$. On the other hand, when correlations are suppressed, i.e., $V=0$ or $\xi_M\leq\xi_S$, a similar analysis for $\rho_I\ll\lambda_L$ provides that the threshold exchange energy is still given by Eq.~\eqref{eq:VortexCriterion}, but with $\xi_M$ replaced by $\xi_S$, under the usual condition $\xi_S\ll\rho_I$. Hence, the first fraction now reads as $D_\phi/D_I$. In either case, with or without magnetic correlations present, the first fraction in the r.h.s. of Eq.~\eqref{eq:VortexCriterion} is expected to be of order one for the systems stu\-died here. Therefore, we focus on the second term which corresponds to the threshold magnetic flux required to pin a vortex, after being converted into an exchange splitting energy via the ensuing magnetoelectric effects. Under the assumption that the first fraction in Eq.~\eqref{eq:VortexCriterion} is of the order of one, we find that in the case of the TI surface states the threshold exchange energy to stabilize a single-unit vortex is $\tilde{I}_z\sim0.9\,{\rm meV}$, while for the Rashba metal is $\tilde{I}_z\sim56\,{\rm meV}$. These results further confirm our earlier conclusions, that is, the TI system is more prominent to exhibit magnetic-island-pinned vortices than the Rashba metal. The difference here is due to the fact that both $\lambda_L$ and ${\cal X}$ for the TI system are about three or four times larger than the respective quantities in the Rashba metal.

A more transparent expression for the above introduced vortex-formation criterion can be obtained by re-expressing the London penetration depth using its definition in terms of the superfluid stiffness which is roughly given as $D\sim\pi E_F/(2\Phi_0^2)$, as obtained from our calculations discussed in Appendix~\ref{app:AppendixGLcoeff}. These manipulations lead to the following formula for a disorder-free system:
\begin{align}
\tilde{I}_z^{\rm clean}=\frac{\ln(\lambda_L/\xi_S)}{\ln\big({\rm max}\{\rho_I,\xi_M\}/{\rm min}\{\rho_I,\xi_M\}\big)}\frac{E_F}{4{\cal X}\Phi_0}\,,
\end{align}

\noi and the following expression in the case that disorder is present:
\begin{align}
\tilde{I}_z^{\rm dirty}=\frac{\ln(\lambda_L/\xi_S)}{\ln\big({\rm max}\{\rho_I,\xi_M\}/{\rm min}\{\rho_I,\xi_M\}\big)}\frac{\pi^2\Delta}{2{\cal X}\Phi_0}\frac{\ell}{\lambda_F}.
\end{align}

\noi From the above expressions and by taking into account that ${\cal X}\Phi_0\sim1$, we conclude that for a clean system the threshold exchange energy induced by the magnetic island on the electrons should be comparable to the Fermi energy in order to stabilize a vortex of a single unit. This is realistic for the TI surface states, since the Fermi ener\-gy is of the order of a few meVs and magnetic-impurity-splittings of this order of magnitude have already experimentally observed in such systems. See for instance Ref.~\onlinecite{KunJiang} and references therein. In contrast, this requirement is very challen\-ging to meet in a Rashba metal in the absence of disorder since the Fermi energy is very large. Within the present framework, the only possible way for a clean Rashba metal to circumvent this obstacle is the presence of strong magnetic correlations. Indeed, the arising renormalization of the spin moment of the magnetic island through the RPA enhancement can provide this way out. Remar\-ka\-bly, however, in the pre\-sen\-ce of substantial disorder the energy scale that controls the vortex pinning is only about one order of magnitude larger than the pairing gap and, thus, typically lies in the low ${\rm meV}$ range. Hence, disorder can strongly facilitate the vortex pinning me\-cha\-nism discussed here even when the Fermi energy is large.

\subsection{Experimental Considerations and Feasibility}

The above discussion considers the threshold exchange energy by assuming that the pairing gap is unaffected by the magnetic island. This is certainly safe to assume when the magnetic exchange energy is much smaller than the pairing gap $\Delta$. However, in the above expressions we found that for clean systems the threshold exchange ener\-gy is required to be of the order of the Fermi ener\-gy which is typically (much) larger than $\Delta$, while for disordered SCs the threshold has to be at least an order of magnitude larger than $\Delta$. Hence, in this work, we are in fact accessing the regime where the magnetic island is expected to significantly in\-fluen\-ce the pai\-ring gap.

From prior studies, it is well-established that a magnetic impurity can suppress the pairing gap locally and even lead to its sign change~\cite{Flatte,Salkola}. On the other hand, since here we are away from the point-like impurity limit and instead exa\-mi\-ne a magnetic island, it is expected that superconductivity will ``melt'' via a first-order phase transition  exactly when the exchange energy reaches the Chandrasekhar-Clogston limit, i.e., $\tilde{I}_z^{\rm cc}=\Delta/\sqrt{2}$~\cite{CClimit}. The latter scenario is possible as long as the conventional pairing term persists and no so-called Fulde-Ferrell-Larkin-Ovchinnikov (FFLO) states become stabilized before reaching the $\tilde{I}_z^{\rm cc}$ value~\cite{FFLO}. Therefore, under the assumption that we exclude the emergence of FFLO phases, the magnetic island leaves the superconducting coherence length unaffected until superconductivity breaks down in the magnetized region at $\tilde{I}_z^{\rm cc}$. When this takes place, the vortex core radius is essentially defined by $\tilde{\rho}_I$. Since in this work we have identified the superconducting coherence length with the vortex core radius, we expect that for exchange energies exceeding $\tilde{I}_z^{\rm cc}$, the magnetic island in this regime effectively behaves as a ``point-like magnetic impurity'' which only influences the normal region inside the vortex core.  Therefore, for strong exchange ener\-gies, $\tilde{\xi}_S\sim\tilde{\rho}_I$ and one of the main assumptions of our theo\-re\-ti\-cal framework breaks down.

From the analysis of the previous section, we found that the threshold exchange energy to stabilize a single-unit vortex for the case of the TI surface states is $\tilde{I}_z\sim0.9\,{\rm meV}$, while for the Rashba metal is $\tilde{I}_z\sim56\,{\rm meV}$. Since in either case the pairing gap is of the order of $1\,{\rm meV}$, it becomes obvious that the superconducting TI surface states appear capable of allowing for the zero-field vorticies proposed in this work, while the superconduc\-ting Rashba metal with a large Fermi energy cannot sustain the superconducting gap for these high exchange energy values that are required to pin a zero-field vortex.

It is important to remark that the above discussion implicitly considered that the pairing gap is of an intrinsic origin, i.e., it adjusts accordingly to minimize the energy of the system. However, considering instead that the quasi-2D SCs of interest inherit a so-called proximity-induced gap  -- due to their coupling to a parent bulk SC -- can provide an escapeway to the above deadend. Indeed, the proximity-induced gap can be sustained for exchange ener\-gies even above $\Delta/\sqrt{2}$, since $\tilde{I}_z^{\rm cc}$ is now eva\-lua\-ted using the pairing gap $\Delta_0$ of the parent SC which is larger than $\Delta$. Noteworthy, such a situation takes place in FeTeSe, where the $\Delta_0$ defines the bulk pai\-ring which is known to be of the order of $2\,{\rm meV}$~\cite{Kreisel_review}, thus rendering the realization of our proposal in this system experimentally feasible. In stark contrast, for a Rashba metal with a Fermi energy of the order of a few hundred ${\rm meVs}$ it is not possible to beat the Chandrasekhar-Clogston limit even when harnessing the superconduc\-ting proximity effect. This is because the pairing order parameter for known superconducting materials is always of the order of a few ${\rm meVs}$. Nonetheless, our mechanism is still applicable to superconducting Rashba systems which feature a Fermi energy of a few ${\rm meVs}$ akin to the TI case discussed earlier. Such a physical scenario can be rea\-li\-zed in Rashba 2DEGs in proximity to conventional or even more preferrably high-T$_{\rm c}$ SCs, see for instance Ref.~\onlinecite{SauPRL}.

\section{Vortex-Majorana Zero Modes}\label{sec:VortexMZMs}

Having identified the conditions under which superconducting vortices can become stabilized by a magnetic island in the absence of an external magnetic field, we now proceed and discuss the topological scenarios that become relevant for each one of the two distinct categories of systems of interest. As a disclaimer, we remark that stabilizing a superconducting vortex does not ne\-ces\-sa\-ri\-ly imply the emergence of vortex-MZMs.

\subsection{Topological Criterion for Vortex-MZMs}\label{sec:SectionIII}

We first introduce the criterion that allows us to infer the conditions under which vortex-MZMs become accessible upon the pinning of a superconducting vortex. Since we are pri\-ma\-ri\-ly interested in modes of a topological origin, we facilitate the discussion by adopting an adiabatic picture, in which the ensuing BdG Hamiltonian in Eq.~\eqref{eq:BdGHamiltonian} can be considered to vary smoothly in terms of the coordinates $\bm{r}=(\rho,\theta)$, which implies that the ope\-ra\-tors $\hat{\bm{p}}$ and $\bm{r}$ commute. This assumption holds as long as the spatial variations of the various fields are ``slower'' than the Fermi wavelength of the system. For our analysis we also discard $A_{x,y}(\bm{r})$ from the gauge invariant momentum $\hat{\bm{\pi}}$ in Eq.~\eqref{eq:BdGHamiltonian}. In addition, we adopt the effective exchange field picture discussed in Sec.~\ref{sec:effectivePicture}, within which the electrons feel the renormalized spin moment field $\tilde{I}_z(\bm{r})\equiv\tilde{I}_z(\rho)=\tilde{I}_zK_0(\rho/\tilde{\rho}_I)$ with the energy scale $\tilde{I}_z=\tilde{S}_z/(2\pi\tilde{\rho}_I^2)$. Under these assumptions the resul\-ting adiabatic BdG Hamiltonian $\hat{\cal H}(\bm{p},\rho,\theta)$ takes the form:
\bea
\hat{\cal H}(\bm{p},\rho,\theta)&=&e^{-i\nu_\phi\theta\tau_z/2}\bigg\{\tau_z\left[\frac{\bm{p}^2}{2m}-E_F+\upsilon_R\big(\bm{p}\times\bm{\sigma}\big)\cdot\hat{\bm{z}}\right]\no\\
&&+\Delta(\rho)\tau_x-\tilde{I}_z(\rho)\sigma_z\bigg\}e^{i\nu_\phi\theta\tau_z/2}\,.
\label{eq:Hadiab}
\eea

\noi To obtain above form, we assumed that the modulus of the superconducting gap $\Delta(\rho)$ and the effective exchange field $\tilde{I}_z(\rho)$ depend only on the radial coordinate. The BdG Hamiltonian belongs to symmetry class D~\cite{Altland,Ryu}, since it possesses an anti-unitary charge-conjugation symmetry $\Xi$, so that the following relation holds $U_\Xi^\dag\hat{\cal H}^*(\bm{p},\rho,\theta)U_\Xi=-\hat{\cal H}(-\bm{p},\rho,\theta)$ with $U_\Xi=\tau_y\sigma_y$. In order to proceed with our analysis of the topological properties of the system, it is convenient to consider that $\Delta(\rho)$ and $M_z(\rho)$ have the simplified piecewise radial dependences $\Delta(\rho)=\Delta\Theta(\rho-\xi_S)$ and $\tilde{I}_z(\rho)=2\tilde{I}_z\Theta(\tilde{\rho}_I-\rho)$~\cite{CommentOnDiskProfile}. $\Theta$ is the Heaviside unit step function.

\subsection{Superconducting Rashba Metal}

The presence of the Schr\"odinger kinetic energy term $\bm{p}^2/2m$ renders the adiabatic BdG Hamiltonian compacti\-fia\-ble in momentum space $(p_x,p_y)\in\mathbb{R}^2$, so that $\mathbb{R}^2$ becomes equivalent to an $\mathbb{S}^2$ sphere. For such a compactified BdG Hamiltonian which belongs to class D, the to\-po\-lo\-gi\-cal invariant which predicts the emergence of vortex-MZMs is of the $\mathbb{Z}_2$ type~\cite{VolovikBook,TeoKane}. As pre\-viously discussed, see for instance Ref.~\onlinecite{TeoKane}, the $\mathbb{Z}_2$ invariant is identified with ${\rm Exp}(i\pi\nu_\phi{\cal C}_1)$, where ${\cal C}_1$ corresponds to the 1st Chern number of the occupied bulk bands of the Hamiltonian in Eq.~\eqref{eq:Hadiab}, calculated in the absence of the vortex, i.e., by setting $\nu_\phi=0$. This implies that the $\mathbb{Z}_2$ invariant is tri\-vial when it is equal to $1$. Hence, when the product $\nu_\phi{\cal C}_1$ is odd one obtains a nontrivial $\mathbb{Z}_2$ invariant equal to $-1$, which predicts the emergence of a single MZM trapped in the core of the supercon\-duc\-ting vortex. This can be viewed as a result of a phase transition from a normal system defined in the radial interval $\rho\in[0,\xi_S]$, to a topological SC living in $\rho\in[\xi_S,\tilde{\rho}_I]$. Notably, a topological SC with $|{\cal C}_1|=1$ becomes stabilized in $\rho\in[\xi_S,\tilde{\rho}_I]$ when the condition $2|\tilde{I}_z|>\sqrt{E_F^2+\Delta^2}$ is satisfied. Hence, when $\nu_\phi$ is odd, a single MZM becomes pinned in the vicinity of the vortex core edge $\rho\sim\xi_S$, thus, extending along the circumference of the boun\-da\-ry determined by $\rho=\xi_S$.

\begin{figure}[t!]
\begin{center}
\includegraphics[width=\columnwidth]{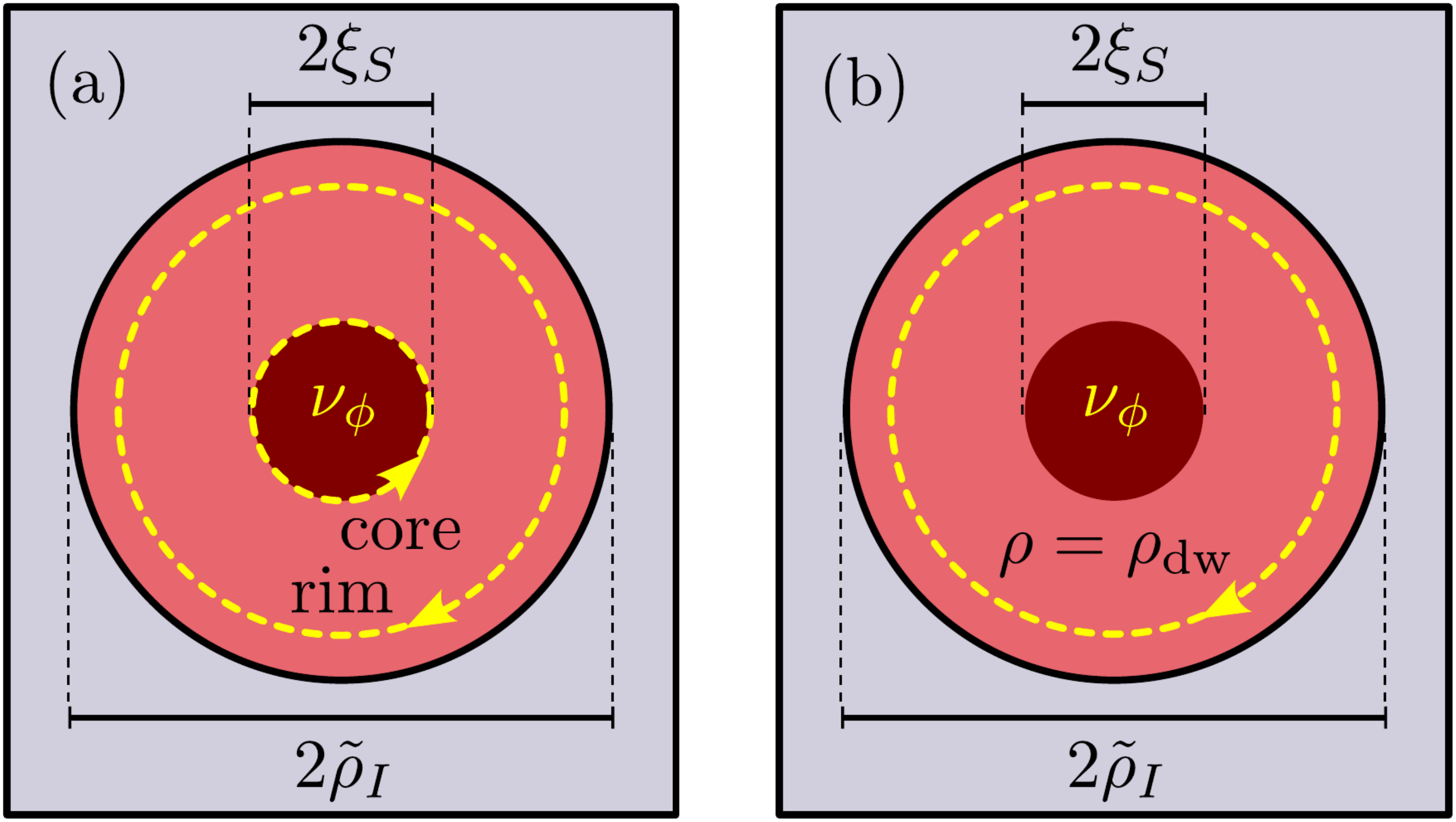}
\end{center}
\caption{Top view of the system and vortex-MZMs for (a) a Rashba SC and (b) superconducting TI surface states. The MZMs are shown with dashed lines and extend uniformly along these. In (a) a pair of core-rim MZMs emerge due to the topological SC realized in the region enclosed. Instead, only a single domain-wall-MZM appears in (b). The arrows indicate the associated Majorana chiral edge modes of the MZMs.}
\label{fig:Figure5}
\end{figure}

Together with the core vortex-MZM, an additional MZM appears at the rim of the vortex located at $\rho=\tilde{\rho}_I$, which is defined as the radial position at which the effective exchange field induced by the magnetic island va\-ni\-shes and a topologically trivial superconducting domain emerges for $\rho>\tilde{\rho}_I$. As pointed out in Ref.~\onlinecite{Alicea}, the emergence of a pair of core-rim vortex-MZMs can be understood as the aftermath of the appearance of dispersive chiral Majorana edge modes. These are located at $\rho=\xi_S$ and $\rho=\tilde{\rho}_I$ and propagate in opposite directions. These appear as a consequence of the bulk nontrivial topology in the domain $\rho\in(\xi_S,\tilde{\rho}_I)$ with $|{\cal C}_1|=1$. For a circum\-fe\-ren\-ce of a finite length and in the absence of a vortex, the chiral Majorana edge modes do not include any zero energy modes. Zero-energy pinning occurs only after a vortex with an odd number of vorticity quanta is introduced, since this twists the electronic wavefunction~\cite{ReadGreen}.

From the above analysis, we infer that a pair of core-rim vortex-MZMs can become stabilized by a magnetic island in the absence of an external magnetic field, thus opening the door to implement the proposal of Ref.~\onlinecite{SauPRL} in a self-tuned fashion. See Fig.~\ref{fig:Figure5}(a) for a schematic depiction. Indeed, the situation described above appears feasible to realize in hybrid SC-semiconductor platforms, where the Fermi energy $E_F$ of the semiconductor can be in principle tuned via gating~\cite{CommentSmallFermiEnergyRashba}. This allows for the criterion $2|\tilde{I}_z|>\sqrt{E_F^2+\Delta^2}$ to be met already for $\tilde{I}_z\sim\Delta/2$. Assuming that the parent SC mediating the superconducting proximity effect features a pairing gap $\Delta_0$ with $\Delta_0>\Delta$, this leads to a window for values of $\tilde{I}_z$ for which both superconductivity in the parent SC and a topological SC in the proximitized system can coexist.

Instead, for heterostructures based on elemental SCs, such as the Pb platforms which have been experimentally addressed in Refs.~\onlinecite{Cren,Menard}, the Fermi energy is very large. In particular, for the latter systems the Fermi ener\-gy is predicted to be about $660\,{\rm meV}$~\cite{PbFermiEnergy}, while other typical metallic SCs are cha\-rac\-te\-ri\-zed by a Fermi ener\-gy which is of the order of a few eVs. Therefore, in these systems, it appears challen\-ging to compensate the Fermi ener\-gy by the scale $|\tilde{I}_z|$. At least, this is in the absence of magnetic correlations. Indeed, our preceding analysis illustrates that the effective spin moment $\tilde{S}_z$ can be significantly enhanced depending on how close is the system to a magnetic instability. As we discussed in Sec.~\ref{sec:effectivePicture}, in the limit of small ${\cal G}$ we have $\tilde{S}_z\approx S_z/\big(1-V\chi_\perp^{\rm spin}\big)$, which shows that the effective spin moment is enhanced by the RPA factor $1/(1-V\chi_\perp^{\rm spin})$. Therefore, we conclude that Rashba SCs with substantial magnetic correlations can boost the exchange field generated by the magnetic island and allow for the system to meet the topological criterion. First principles calculations in Pb systems deposited on top of Si(111) surfaces indicate that an onsite Hubbard interaction of the order of $1\,{\rm eV}$ becomes re\-le\-vant~\cite{PbHubbard}, thus, implying that magnetic correlations are present. However, for such high values of $\tilde{I}_z$ superconductivity cannot be sustained in the region where the SC couples to the magnetic island. In such a case, we expect the coherence length to become equal to $\tilde{\rho}_I$, hence, not allowing for topological superconductivity to develop.

\subsection{Superconducting TI Surface States}

To describe the TI surface states, it is eligible to first take the limit $m\rightarrow\infty$, which essentially eliminates the Schr\"odinger kinetic ener\-gy term from the Hamiltonian in Eq.~\eqref{eq:Hadiab}. It is important to remark that a Schr\"odinger kinetic ener\-gy remains a legitimate term also here, as long as it takes  values that do not lead to more than a single topologically-protected helical branch within the energy window in which helical surface states emerge. As a matter of fact, the possibility to have a well-defined model which is linear in momenta and, thus,  non-compactifiable in momentum space, directly reflects that we are here dealing with a boundary rather than a bulk model Hamiltonian. This is an important difference compared to the bulk superconducting Rashba system discussed in the previous section, since in the present case the impossibi\-li\-ty to compactify the momentum space further implies that any topological invariant that can be defined can at most be fractio\-nal\-ly quantized~\cite{VolovikBook,Ryu}.

The above aspect becomes transparent in the case $E_F=0$, where the adiabatic Hamiltonian defined for $\nu_\phi=0$ becomes block dia\-go\-nal due to the emergence of a unitary symmetry $\big[\hat{\cal H}(\bm{p},\rho,\theta;\nu_\phi=E_F=0),\tau_x\sigma_z\big]=0$. We transfer to a frame in which the unitary symmetry operator $\tau_x\sigma_z$ is block-diagonal by considering the unitary transformation, ${\cal U}^\dag\hat{\cal H}(\bm{p},\rho,\theta;\nu_\phi=E_F=0){\cal U}$, where ${\cal U}=(\tau_x\sigma_z+\tau_z)/\sqrt{2}$, and obtain the following two block Hamiltonians $
 \hat{\cal H}_\tau(\bm{p},\rho,\theta;\nu_\phi=E_F=0)=\bm{g}_\tau(\bm{p},\rho)\cdot\bm{\sigma}$. Here, we set $\bm{g}_\tau(\bm{p},\rho)=\tau\big(-\upsilon_Rp_y,\upsilon_Rp_x,\Delta(\rho)-\tau \tilde{I}_z(\rho)\big)$,  where the quantum number $\tau=\pm1$ represents the unitary symmetry operator in this block diagonal space and, thus, labels the BdG Hamiltonians of the two blocks. By means of straightforward calculations we find that each block has a fractional adiabatic Chern number ${\cal C}_{1,\tau}(\rho)$ defined at a given value of $\rho$, which is given by the expression ${\cal C}_{1,\tau}(\rho)={\rm sgn}\big[\tau\Delta(\rho)-\tilde{I}_z(\rho)\big]/2$. Evidently, the two invariants are fractional, their signs are generally different, and depend on the relative strength of the pairing gap and the exchange splitting. Hence, on such a topological surface, Majorana excitations can be trapped only at domain walls, across which the energy scale hie\-rar\-chy of $|\Delta(\rho)|$ and $|\tilde{I}_z(\rho)|$ is inverted, since this will trigger a gap closing in one of the adiabatic block Hamiltonians.

 From the above analysis, we conclude that for superconducting TI surface states, the topological criterion has to be accordingly modified since MZMs can be trapped only when variations $\delta{\cal C}_1$ of the 1st Chern  number ${\cal C}_1$ take place. Hence, the ensuing $\mathbb{Z}_2$ invariant in the present case is instead ${\rm Exp}(i\pi\nu_\phi\delta{\cal C}_1)$, and indicates the emergence of a single MZM trapped at a domain wall. This $\mathbb{Z}_2$ quantity allows us to establish an index theorem, in analogy to the ce\-le\-bra\-ted Atiyah-Singer theorem~\cite{Atiyah}, which predicts the zero modes which arise at a mass domain wall of a Dirac electron~\cite{JackiwRebbi}. Here, the spatially-dependent Dirac mass which features the domain wall is identified with the term $\tau\Delta(\rho)-\tilde{I}_z(\rho)$ of each block adiabatic Hamiltonian.

A domain wall located at $\rho=\rho_{\rm dw}$ traps a single dispersive chiral Majorana mode with eigenenergies $E_n=n\hbar\upsilon_R/\rho_{\rm dw}$~\cite{ReadGreen,Volovik99,Alicea}, with $n\in\mathbb{Z}$. This expression holds for low energies and is obtained by assuming that $\rho_{\rm dw}$ is sufficiently large to safely allow us to discard any curvature effects. The arising domain-wall-MZM is analogous to the rim MZM obtained for the case of the Rashba SC studied in the previous section. That is, it corresponds to the $n=0$ mode of the respective chiral Majorana mode sequence and extends uniformly along the circumference of the domain wall. See Fig.~\ref{fig:Figure5}(b) for a schematic depiction. Note that for small values of $\rho_{\rm dw}$, the nonzero energy modes ($n\neq0$) may be pushed above the bulk energy gap and end up to be unob\-ser\-va\-ble, thus lea\-ving the MZM as the only in-gap excitation at the vortex core.

We thus conclude that such a ``mass'' domain wall traps a vortex-MZM as long as the above criteria are fulfilled, i.e., $\Delta(\rho)$ is required to vanish at least at a single point within the area covered by the magnetic island, so that a nonzero vorticity can be defined. The re\-co\-ve\-ry of the ``bulk'' value for the pairing gap is expected to occur within a lengthscale given by the coherence length $\xi_S$. However, the sign change of $\tilde{I}_z(\rho)-\Delta(\rho)$, which determines the location of the vortex MZM, can take place at a radius which is ge\-ne\-ral\-ly unrelated to $\xi_S$, and its location depends on the properties of the exchange field $\tilde{I}_z(\rho)$. For clarity, in Appendix~\ref{app:AppendixMZM} we present a detailed analysis of the emergence of the vortex-MZM in the case of superconducting TI surface states in an exchange field.

\section{Conclusions and outlook}
\label{sec:Conclusions}

In this work, we theoretically propose and investigate a mechanism to stabilize superconducting vortices in a quasi-2D Rashba SC in the absence of an applied magnetic field. In our approach, the required magnetic flux to pin a vortex is instead provided by a magnetic island, i.e., a spatially-extended magnetic impurity. The key ingre\-dients and assumptions of our proposal are: (i) the spatial extent of the magnetic island dictated by the lengthscale $\rho_I$ is assumed to greatly exceed the superconducting coherence length $\xi_S$ and the Fermi wavelength $\lambda_F$ defined in the normal phase of the SC, (ii) the magnetic island couples to the electrons of the SC only via an exchange coupling, (iii) the spin moment of the island is con\-si\-de\-red to be oriented out-of-the-plane due to crystal field effects, (iv) the magnetic island does not introduce Yu-Shiba-Rusinov states~\cite{CommentOnYSR}, (v) the spin moment carried by the island is converted into magnetic flux by means of Zeeman and Rashba magnetoelectic effects, and (vi) we take into account the possible pre\-sence of magnetic correlations in the SC which induce an electronic magnetization. To address the problem, we employ a phe\-no\-me\-no\-lo\-gi\-cal GL functional which is motivated by and derived from a representative microscopic model Hamiltonian. In addition, our GL for\-ma\-lism is equipped with a concrete spatial profile for the spin moment field of the island, which allows us to derive closed form analytical expressions for the magnetic field and the induced magnetization characterizing the vortex.

With the vortex solution at hand, we introduce two important criteria. The first identifies the conditions that should be met in order for zero-field vortices to appear. For a clean system, we conclude that a magnetic-island-induced vortex with a single unit of magnetic flux is stabilized when the ``dressed'' exchange ener\-gy felt by the electrons due to the island becomes comparable to the Fermi energy characterizing the normal phase of the SC. In contrast, in the case of a strongly disordered system, the required threshold exchange energy is found to be about an order of magnitude larger than the pairing gap of the SC. Hence, this remar\-ka\-ble modification offers a loophole for our mechanism to be realized even in SCs with a large Fermi energy. The second criterion that we obtain concerns the subsequent trapping of MZMs by the vortex pinned by our nonstandard mechanism. For an effective spin moment field of the magnetic island which varies in space slower than the Fermi wavelength of the SC, we find that MZMs can be stabilized as long as the effective exchange energy felt by the electrons in the SC primarily exceeds the Fermi energy of the system.

We find that since a Rashba SC effectively behaves as a bulk p+ip topological SC, pairs of MZMs appear for a super\-con\-duc\-ting vortex of an odd vor\-ti\-ci\-ty. This is ana\-lo\-gous to what has been discussed for bulk spinless chiral p+ip SCs~\cite{ReadGreen,Volovik99,Alicea}, as well as for conventional SCs with spin-orbit~\cite{Cren} or magnetic skyrmion defects~\cite{Mirlin,Mesaros,Garnier}. Spe\-ci\-fi\-cal\-ly, one MZM of the pair is trapped at the vortex core, while an additional MZM appears at the outer boundary of the system. The latter is either associated with the termination boun\-dary of the 2D p+ip SC, or, with the termination of the defect. Essentially, the MZM pair in these cases resembles the pair of MZMs stemming from the two terminations edges of 1D to\-po\-lo\-gi\-cal nanowires~\cite{KitaevUnpaired,LutchynPRL,OregPRL}. In the case of superconducting TI surface states, however, a single vortex binds only a single domain-wall-MZM whose radial location or ``orbit'' is determined from the compensation of the magnetic exchange and pairing gaps (assuming charge neutrality). Nonetheless, the MZMs still have to come in pairs. These pairs either ori\-gi\-na\-te from an even number of vortices confined at a given TI surface, or, from the two ends of vortex lines piercing two opposite TI surfaces, akin to the multi-vortex scenarios earlier proposed in Ref.~\onlinecite{KunJiang}.

Our findings have important implications for various experiments. First of all, our results show that for clean or disordered systems with a small Fermi energy zero-field vortices and MZMs go hand in hand. In contrast, for disordered SCs with large Fermi energies which allow for zero-field vortices, concomitant MZMs become accessible only for substantially large exchange energies. Our ana\-ly\-sis illustrates that a possible way to achieve this is by relying on the presence of magnetic correlations. Indeed, for correlated Rashba SCs the effective exchange energy felt by the electrons increases upon tuning the system closer to a magnetic instability. In this case, the exchange energy becomes renormalized by an RPA factor which effectively boosts the spin moment of the island. However, as we pointed out in the main text, in Rashba SCs with interaction-driven intrinsic superconductivity,  exchange energies much larger than the pairing gap may suppress superconductivity in the magnetized region and can prove detrimental for the emergence of MZMs.

Our proposal and study is motivated by recent expe\-ri\-ments in (i) FeTeSe iron-based SCs~\cite{JiaXinYin,QAHVexpEarly,QAHVexp} and (ii) in Pb systems deposited on top of Si(111) under the influence of magnetic islands~\cite{Cren,Menard}. When it comes to FeTeSe systems a zero-field so-called quantum anomalous vortex has been theoretically proposed~\cite{KunJiang} and experimentally observed~\cite{QAHVexpEarly,QAHVexp} in the case of Fe adatoms. These act as magnetic impurities in the antipodal limit than the one studied here. Indeed, such impurities have a cha\-rac\-te\-ri\-stic lengthscale $\rho_I$ which is smaller or similar to $\xi_S$. Despite this crucial difference, our prediction of a zero-field vortex in the regime $\lambda_L\gg\rho_I\gg\xi_S$, where $\lambda_L$ is the London penetration depth for the SC, is in overall agreement with the analysis of Ref.~\onlinecite{KunJiang}. In both regimes, it is the smallness of the Fermi energy that guarantees the simultaneous pinning of zero-field vortices and their partner MZMs. Although the requirements are similar, the two mechanisms are distinct. In Ref.~\onlinecite{KunJiang} the presence of Yu-Shiba-Rusinov states are important for the stabilization of vortices and the emergence of MZMs. Here, instead, such in-gap states are not relevant, due to the extended nature of the magnetic element, which renders the emergence of zero-field vortices as a phenomenon which is tied to the bulk electrons of the SC~\cite{CommentOnYSR}. Our analysis predicts that composite vortex-Majorana excitations can be pinned in iron-based SCs by magnetic islands. From an experimental point of view, realizing a zero-field vortex in these systems without the involvement of Yu-Shiba-Rusinov states may provide a more solid basis for understanding such effects, while at the same time it can provide a fertile ground for a stronger degree of manipulability of MZMs. At the same time, the smallness of the Fermi energy of the surface states of FeTeSe provides the unique opportunity to distinguish MZMs from Caroli - de Gennes - Matricon states, since these have an experimentally-resolvable energy splitting. As a matter of fact, an alternative way to pin such zero-field vortices in FeTeSe systems may be useful for resolving uncertainties that have arisen concerning the experimental interpretation of MZMs in FeTeSe~\cite{Milan,Freek}.

Moving on, in the experiments studied in Refs.~\onlinecite{Cren,Menard} the magnetic islands coupled to Pb have a spatial extent which is also much smaller or comparable to $\xi_S$. Our analysis can predict whether our proposal is applicable to these systems when $\rho_I\gg\xi_S$. The presence of strong disorder and the absence of Yu-Shiba-Rusinov states are both advantageous. However, trapping vortex-MZMs in these systems appears challenging. This is due to the large Fermi energy that typically characterizes elemental SCs and demands accordingly large exchange energies to pin zero-field vortices. Eventhough magnetic correlations present in these systems can effectively enhance the exchange energy, the expected unsustainability of superconductivity in the magnetized region strongly disfavors the implementation of our proposal in these systems.

A prominent alternative that allows to circumvent the above hindrances already from the outset, is to consider ferromagnet-semiconductor-SC hybrids~\cite{SauPRL}. In these systems, the vortex and MZMs are expected to be pinned in the semiconductor which experiences a proximity-induced pairing. Thanks to the gate-tunability of the semiconductor, its Fermi energy can be controllably set to be small, therefore enabling this system to simultaneously exhibit zero-field vortices along with vortex-MZMs.

We conclude this work by noting that the cooperative interplay between vortices and magnetic islands has previously been studied in the context of various correlated SCs~\cite{FischerRMP}. For example, in the field of cuprate SCs, the induction of magnetic regions nucleated by  vortex cores has been extensively discussed in the literature~\cite{Arovas,AndersenSO5,Chen2002,Ghosal2002,Zhu2002,Takigawa2003,Udby,Schmid}. These studies all refer to Abrikosov vortices induced by externally applied magnetic fields, and are therefore distinct from the fin\-dings of the current paper centered on the criteria for the emergence of zero-field superconduc\-ting vortices by magnetic islands.

\section*{Acknowledgements}

We thank M. Roig and H.~O.~M. Sura for prior motivating collaborations and discussions during the project.

\appendix

\section{Useful Relations and Integrals}\label{app:AppendixA}

In this appendix, we provide a number of useful expressions and results that we employed in our analysis of the vortex solution. Specifically, in order to ``invert'' the Fourier and Hankel transforms, we decomposed the va\-rious terms using the following identities:
\bea
\frac{1}{\big(q^2+\frac{1}{a^2}\big)\big(q^2+\frac{1}{b^2}\big)}=\frac{(ab)^2}{a^2-b^2}\left(\frac{1}{q^2+\frac{1}{a^2}}-\frac{1}{q^2+\frac{1}{b^2}}\right),\no\\\\
\frac{q^2+\frac{1}{c^2}}{\big(q^2+\frac{1}{a^2}\big)\big(q^2+\frac{1}{b^2}\big)}=\frac{(ab)^2}{a^2-b^2}\left(\frac{\frac{1}{c^2}-\frac{1}{a^2}}{q^2+\frac{1}{a^2}}-\frac{\frac{1}{c^2}-\frac{1}{b^2}}{q^2+\frac{1}{b^2}}\right).\no\\
\eea

In order to infer the vortex stability, we evaluated the energy of the vortex ground state with the approximate forms of the following two exact results:
\bea
&&\int_{\xi_S}^\infty d\rho\,\rho\, K_0^2(\rho/a)=\frac{\xi_S^2}{2}\Big[K_1^2(\xi_S/a)-K_0^2(\xi_S/a)\Big]\,,\\
&&\int_{\xi_S}^\infty d\rho\,\rho\, K_0(\rho/a)K_0(\rho/b)=\frac{(ab)^2}{a^2-b^2}\times\no\\
&&\left[K_0(\xi_S/a)K_1(\xi_S/b)\frac{\xi_S}{b}-K_0(\xi_S/b)K_1(\xi_S/a)\frac{\xi_S}{a}\right],\qquad\,
\eea
\noi with $a$ and $b$ corresponding to two positive and unequal variables. Here, $K_1(z)$ denotes the first order modified
Bessel function of the second kind with $z\in[0,\infty)$, which is also related to $K_0(z)$ through $K_1(z)=-dK_0(z)/dz$.

\section{Magnetic Island with a Disk Profile}\label{app:AppendixB}

In this section we derive the spin-to-vorticity conversion coefficient $\zeta$ in the case of a magnetic island dictated by a disk-like spin-moment spatial profile of the form $I_z(\rho)=\big(S_z/\pi\rho_I^2\big)\Theta(\rho_I-\rho)$, as explained in Ref.~\onlinecite{CommentOnDiskProfile}. After Eq.~\eqref{eq:HzFourier0}, the magnetic field is found by inverting the following expression:
\bea
\frac{H_z(\bm{q})}{\Phi_0}=\frac{2J_1(q\rho_I)}{q\rho_I}\frac{\nu_I}{(q\lambda_L)^2+1}+\frac{\nu_\phi}{(q\lambda_L)^2+1}\,.
\eea

\noi While the first contribution is not straightforward to be inverted in the general case, such a procedure becomes simplified in the extreme cases $\rho_I\gg\lambda_L$ and $\rho_I\ll\lambda_L$. In either case, the first contribution is governed by the term which contains the dominant length scale.

\subsubsection{Case $\rho_I\gg\lambda_L$}

According to the above mentioned ``recipe'', we have:
\bea
\frac{H_z(\bm{q})}{\Phi_0}\simeq\frac{2J_1(q\rho_I)}{q\rho_I}\nu_I+\frac{\nu_\phi}{(q\lambda_L)^2+1}\,,
\eea

\noi which leads to the real space fields:
\bea
\frac{H_z(\rho)}{H_0}&=&\left(\frac{\lambda_L}{\rho_I}\right)^22\nu_I\Theta(\rho_I-\rho)+\nu_\phi K_0(\rho/\lambda_L),\quad\\
B_z(\rho)&=&H_0\nu_\phi K_0(\rho/\lambda_L).
\eea

\noi Straightforward manipulations analogous to the ones in the main text yield that:
\begin{align}
\zeta_{\rho_I\gg\lambda_L}=-\frac{1}{D_\phi}\left(\frac{\lambda_L}{\rho_I}\right)^2,
\end{align}

\noi which becomes identical to the expression presented in Eq.~\eqref{eq:Limitrholargerlambda} after dropping the unity in the prefactor ente\-ring the latter. This approximation is well-justified for substantially large values of $\rho_I/\lambda_L$. From the above, we conclude that the shape of the spatial profile mainly mo\-di\-fies the slowly-varying prefactors in $\zeta$, while it leaves the characteristic $(\lambda_L/\rho_I)^2$ dependence unaffected.

\subsubsection{Case $\rho_I\ll\lambda_L$}

In this limit, the magnetic field becomes simplified and obtains the approximate form:
\bea
\frac{H_z(\bm{q})}{\Phi_0}\simeq\frac{\nu_I+\nu_\phi}{(q\lambda_L)^2+1}\,,
\eea

\noi which leads to the real space fields:
\bea
H_z(\rho)&=&H_0(\nu_I+\nu_\phi) K_0(\rho/\lambda_L),\quad\\
\frac{B_z(\rho)}{H_0}&=&(\nu_I+\nu_\phi) K_0(\rho/\lambda_L)-\left(\frac{\lambda_L}{\rho_I}\right)^22\nu_I\Theta(\rho_I-\rho).\no\\
\eea

\noi Using the above results, we directly find the conversion coefficient:
\begin{align}
\zeta_{\rho_I\ll\lambda_L}=\frac{1}{D_\phi}\left(\frac{\lambda_L}{\rho_I}\right)^2,
\end{align}

\noi which is identical to the one in Eq.~\eqref{eq:Limitrhosmallerlambda}, if we set $D_I=2$.

\section{Weak vs Strong Coupling Regime}\label{app:AppendixStrong2Weak}

As we emphasized in Sec.~\ref{sec:Spin2Flux}, it is helpful to understand our results concerning the stability of the vortex ground state in certain li\-mits of interest. Specifically, it is desired to study the weak and strong coupling regimes governing the mixing of the magnetic and magnetization fields. In the weak coupling regime $|{\cal G}|\ll|{\cal G}|_{\rm max}$ and $\rho_\pm\approx\{\xi_M,\lambda_L\}$, while for strong couplings $|{\cal G}|$ becomes equal to $|{\cal G}|_{\rm max}$, thus resul\-ting in $\rho_+=\rho_-$. In either scenario, it is inte\-re\-sting to study the outcomes for the various hierarchies between the values of $\xi_M$, $\lambda_L$, and $\rho_I$.

\subsection{Weak Coupling Limit}

In the case that $|{\cal G}|$ is sufficiently weak, we consider the spin-to-vorticity conversion factor $\tilde{\zeta}$ at zeroth order with respect to ${\cal G}$. This further implies that at this level of approximation the variables $\rho_\pm$ enter at zeroth order in ${\cal G}$. Hence, in the weak coupling limit, the lengthscales $\rho_\pm$ are approximately given by $\lambda_L$ and $\xi_M$. Due to the fact that $\rho_-\geq\rho_+$, we obtain the correspondence $\rho_-={\rm max}\{\lambda_L,\xi_M\}$ and $\rho_+={\rm min}\{\lambda_L,\xi_M\}$. In the following, we examine weak-coupling-limit possibilities by varying the hierarchies holding for the lengths $\xi_M$, $\lambda_L$, and $\rho_I$.

\subsubsection*{Cases:\quad $\lambda_L\ll\xi_M\ll\rho_I$ and $\lambda_L\ll\rho_I\ll\xi_M$}

In the event that the London penetration depth is the smallest lengthscale out of the three, we find that both si\-tua\-tions can be described compactly in terms of the formula:
\begin{align}
\tilde{\zeta}\approx\frac{{\rm min}\{D_I,D_M\}-{\rm max}\{D_I,D_M\}}{2D_\phi}\left(\frac{\lambda_L}{{\rm max}\{\rho_I,\xi_M\}}\right)^2,  
\end{align}

\noi where we defined the stiffness $D_M=\ln\big(\xi_M/\xi_S\big)$. We observe that the structure of the above result strongly resembles the one obtained in Eq.~\eqref{eq:Limitrholargerlambda}. Notably, we find that when $\xi_M\gg\rho_I$ the vorticity formation is dictated by the magnetic correlation length rather than the radius of the magnetic island. We benchmark the above appro\-xi\-ma\-tion by choosing the same parameter values used in Fig.~\ref{fig:Figure3}(a)-(b). Our approximation yields $\tilde{\zeta}\simeq0.07\times 10^{-2}$ which is close to the exact value $\tilde{\zeta}=0.08\times 10^{-2}$.

\subsubsection*{Cases: $\xi_M\ll\rho_I\ll\lambda_L$ and $\rho_I\ll\xi_M\ll\lambda_L$}

We now consider the antipodal limit, in which the London penetration depth corresponds to the largest lengthscale. Once again, both possible scenarios can be compactly expressed by employing a single formula:
\begin{align}
\tilde{\zeta}\approx\frac{{\rm max}\{D_I,D_M\}-{\rm min}\{D_I,D_M\}}{2D_\phi}\left(\frac{\lambda_L}{{\rm max}\{\rho_I,\xi_M\}}\right)^2,\label{eq:ZetaApprox}
\end{align}

\noi which is identical - up to an overall sign - to the one obtained in the previous paragraph. This expression is also analogous to Eq.~\eqref{eq:Limitrhosmallerlambda}. Compared to the outcome $\tilde{\zeta}=34\times10^{-2}$ obtained for $|{\cal G}|=0$ in Fig.~\ref{fig:Figure3}(c)-(d), the here-derived approximate formula slightly overestimates the exact result since it provides $\tilde{\zeta}\simeq44\times10^{-2}$.

\subsubsection*{Cases: $\xi_M\ll\lambda_L\ll\rho_I$ and $\rho_I\ll\lambda_L\ll\xi_M$}

So far we considered scenarios in which the London penetration depth was the largest or the smallest lengthscale out of the three. Now, we address the two remaining cases where $\lambda_L$ lies in the middle of the hierarchy of the three quantities of interest. By considering this situation, we find that the conversion factor approximately reads as:
\begin{align}
\tilde{\zeta}\approx\left(1-\frac{D_I+D_M}{2D_\phi}\right)\left(\frac{\lambda_L}{{\rm max}\{\rho_I,\xi_M\}}\right)^2\,,
\end{align}

\noi The above expression is quite similar to the one in Eq.~\eqref{eq:Limitrholargerlambda}, with the difference that the lengthscale which dictates the properties of the magnetic island is given by the ${\rm max}\{\rho_I,\xi_M\}$. In fact, this trend was observed in all the scenarios treated here within the weak coupling limit. Compared to the exact result $\tilde{\zeta}=1.12\times10^{-2}$ obtained in the cases shown in Fig.~\ref{fig:Figure3}(e)-(f), our approximation overestimates the exact value since it gives $\tilde{\zeta}\simeq1.73\times10^{-2}$.

\subsection{Strong Coupling Limit}

Having examined the weak coupling regime in detail, we now explore the other extreme limit, i.e., the one in which the coupling between magnetization and magnetic fields is the strongest possible. Thus, here we have the condition $|{\cal G}|=|{\cal G}|_{\rm max}$, which leads to $\rho_\pm\rightarrow\bar{\rho}$ with the common lengthscale $
1/\bar{\rho}=\sqrt{\big(1/\xi_M^2+1/\lambda_L^2\big)/2}$. The arising coincidence of $\rho_\pm$ leads to further simplifications. In the same spirit of the previous paragraph, also here we focus on all the possible hierarchies for $\{\xi_M,\lambda_L,\rho_I\}$.

\subsubsection*{Cases:\quad $\lambda_L\ll\xi_M\ll\rho_I$ and $\lambda_L\ll\rho_I\ll\xi_M$}

As it is customary, we begin by considering the case where the London penetration depth is the smallest lengthscale out of the three. We find that the same approximate formula holds for both possibilities, that is:
\begin{align}
\tilde{\zeta}\approx2\left(1-4\frac{D_I-1}{4D_\phi+\ln4-1}\right)\left(\frac{\lambda_L^2}{\xi_M\rho_I}\right)^2\,.
\end{align}

\noi We now benchmark the above approximate formula. For the case depicted in Fig.~\ref{fig:Figure3}(a), we find that exact and approximate results coincide and give $\tilde{\zeta}=0.019\times20^{-2}$. On the other hand, for the case depicted in Fig.~\ref{fig:Figure3}(b) the exact result is $\tilde{\zeta}=0.011\times10^{-2}$, while our approximation yields the slightly smaller value, i.e., $\tilde{\zeta}\simeq 0.008\times10^{-2}$.

\subsubsection*{Cases:\quad $\xi_M\ll\lambda_L\ll\rho_I$ and $\xi_M\ll\rho_I\ll\lambda_L$}

The next cases to examine concern the limit in which the magnetic correlation length is the smallest. Both cases can be approximately described by the following expression in the strong coupling regime:
\begin{align}
\tilde{\zeta}\approx2\left(\frac{\xi_M}{\rho_I}\right)^2\,.
\end{align}

\noi The above result, already reveals a stark deviation from the findings in the weak coupling limit, since now dif\-ferent hierarchies get effectively bunched together.
We test our approach for both scenarios. First we consider the hierarchy in Fig.~\ref{fig:Figure3}(e), for which the actual value is $\tilde{\zeta}=0.745\times10^{-2}$, while the approximate one is $\tilde{\zeta}=2.0\times10^{-2}$.
Therefore, our approach somehow overestimates the precise value, at least for the parameter values chosen here. In a similar fashion, we focus on Fig.~\ref{fig:Figure3}(c), in which the actual value of the conversion factor is $\tilde{\zeta}=5.34\times10^{-2}$ and the approximate one is $\tilde{\zeta}=8\times10^{-2}$. Therefore, a similar trend is observed also here, i.e., our approach seems to predict somehow larger values for the spin-to-flux conversion which, nonetheless, are of the same order of magnitude.

\subsubsection*{Case:\quad $\rho_I\ll\lambda_L\ll\xi_M$}

We now proceed by examining the hierarchy Fig.~\ref{fig:Figure3}(f). In this case, we find the approximate formula:
\begin{align}
\tilde{\zeta}\approx\left[1-\frac{2\big(D_\phi+D_I\big)+\ln2+2}{4D_\phi+\ln4-1}\right]\left(\frac{\lambda_L}{\xi_M}\right)^2\,.
\end{align}

\noi The exact result obtained in Fig.~\ref{fig:Figure3}(f) is $\tilde{\zeta}=1.28\times10^{-2}$, while our approximate method yields a value similar to this, i.e., $\tilde{\zeta}\simeq0.85\times10^{-2}$.

\subsubsection*{Case:\quad $\rho_I\ll\xi_M\ll\lambda_L$}

The last case to be examined corresponds to the scenario shown in Fig.~\ref{fig:Figure3}(d). For this hierarchy, we obtain that the conversion factor approximately reads as:
\begin{align}
\tilde{\zeta}\approx2\big(D_M-D_I\big)+\ln2-1\,.
\end{align}

\noi We find that compared to the actual value $\tilde{\zeta}=32.7\times10^{-2}$ in Fig.~\ref{fig:Figure3}(d), our approximation yields instead $\tilde{\zeta}=291.2\times10^{-2}$. Hence, for these parameter values our approximation deviates substantially from the actual result. This discrepancy is due to the fact that the three lengthscales are not sufficiently separated for this approximation to hold when choosing $\rho_I=150\xi_S$, $\xi_M=750\xi_S$, and $\lambda_L=1500\xi_S$. Indeed, by considering much larger values for $\lambda_L$, we find an improved agreement of the actual and approximate expressions.

\section{Ginzburg-Landau Coefficients}
\label{app:AppendixGLcoeff}

In this appendix, we employ the microscopic model in Eq.~\eqref{eq:Hamiltonian} to evaluate the GL coefficients in the case of superconducting TI surface states and a Rashba metal. Note that this model has been previously considered in Refs.~\onlinecite{PershogubaCurrents,Anatomy} to evaluate $\chi_R$ in different limits. In addition, Ref.~\onlinecite{Anatomy} also obtained the out-of-plane susceptibility $\chi_\perp^{\rm spin}$ for the case of a Rashba metal (TI) when a conventional pai\-ring gap is present (absent). Below, all quantities are expressed in the original SI unit system.

\subsection{Superconducting Topological Surface States}\label{sec:TIstates}

From Ref.~\onlinecite{Anatomy}, one finds that the magnetoelectic coefficient is discontinuous across $\Delta=0$ and that for $\Delta>0$ it takes the form:
\begin{align}
\chi_R=\frac{{\rm sgn}(\mu)}{8\Phi_0}\left[f_\delta+\delta^2\ln\left(\frac{\delta}{1+f_\delta}\right)\right],
\end{align}

\noi which is parametrized using the function $f_\delta=\sqrt{1+\delta^2}$, that depends on the dimensionless variable $\delta=\Delta/E_{\rm soc}$. In the above we defined the Rashba SOC energy as $E_{\rm soc}=\upsilon_R\hbar k_F$. Here, $E_{\rm soc}$ coincides with the Fermi ener\-gy $E_F=|\mu|$. Moreover, it is important to note that $\chi_R$ is independent of the strength of the SOC when $\delta=0$, even though this coefficient is nonzero only in the pre\-sen\-ce of the Rashba SOC. This reflects an underlying quantum anomaly and arises due to the Dirac nature of the Rashba system. See also See Ref.~\onlinecite{Anatomy} for the connection between the magnetoelectric coefficient and topology.

To infer the coefficient for the Zeeman-mediated coupling, it is required to obtain the spatially uniform out-of-spin susceptibility $\chi_\perp^{\rm spin}$, see also Eq.~\eqref{eq:ZeemanConversion}. This quantity has been previously evaluated for the model of Eq.~\eqref{eq:Hamiltonian} in Ref.~\onlinecite{Anatomy} in the absence of superconductivity. Here, we also extend this calculation when a nonzero pairing gap is present. The related technical details are presented in Appendix~\ref{app:AppendixSusce} and lead to the expression:
\begin{align}
\chi_\perp^{\rm spin}
\simeq\frac{1}{2\pi(\upsilon_R\hbar)^2}\left\{\Lambda-E_F\left[f_\delta-\delta^2\ln\left(\frac{\delta}{1+f_\delta}\right)\right]\right\},\label{eq:SpinSusc}
\end{align}

\noi where the approximate character of the above result
stems firstly from assuming that the ultraviolet cutoff ener\-gy scale $\Lambda$ satisfies $\Lambda\gg E_F,\Delta$ and, secondly from accordingly simplifying the expression for the contribution of the high-energy degrees of freedom. In contrast, the term $\propto E_F$ is
exact and constitutes the contribution of the Dirac point. In the remainder, we restrict
to the case $g=2$, and find the following expression:
\begin{align}
\chi_Z
=\frac{1}{4\Phi_0}\left\{\frac{\upsilon_\Lambda}{\upsilon_R}-\frac{\upsilon_F}{\upsilon_R}\left[f_\delta-\delta^2\ln\left(\frac{\delta}{1+f_\delta}\right)\right]\right\}.\label{eq:SpinSusc}
\end{align}

\noi The above was obtained after replacing the Bohr magneton by its defining relation $\mu_B=e\hbar/(2m_e)$, where $m_e$ is the bare electron mass. Moreover, in the above we in-
troduced the energy-cutoff and Fermi velocities according
to $\upsilon_\Lambda=\Lambda/(m_e\upsilon_R)$ and $\upsilon_F=\hbar k_F/m_e$, respectively.

In order to proceed, we evaluate the lengthscales $\xi_M$ and $\lambda_L$. To find $\xi_M$ it is only left to obtain
$c_M$ which corresponds to the magnetic stiffness, since $\alpha_M$
contains the interaction strength $V$, $\chi_\perp^{\rm spin}$, and ${\cal X}$. The coefficient $c_M$ can be read out from the wave vector dependent out-of-plane spin susceptibility $\chi_\perp^{\rm spin}(\bm{q})$. For a derivation and additional details see Appendix~\ref{app:AppendixMagStiff}. Tedious but straightforward calculations yield the formula:
\begin{align}
c_M=\frac{1}{16\pi E_F}\left[\frac{2}{f_\delta}-f_\delta-\delta^2\ln\left(\frac{\delta}{1+f_\delta}\right)\right]\,,
\label{eq:cMTI}
\end{align}

\noi which is indeed positive as anticipated.

The last physical quantity that remains to be evaluated in order to investigate the stabilization of vortices for the system in question is the London penetration depth $\lambda_L$. The evaluation of the superfluid stiffness for superconducting Dirac electrons has already been examined in detail for the case of graphene in the Dirac regime~\cite{KopninSoninPRL,KopninSoninPRB,Liang,WangAssiliPKletter,WangAssiliPKarticle,WangPK}. The calculation here is similar, because the Rashba SOC is mapped to the orbital-sublattice coupling found in graphene. The only essential difference compared to the prior study in graphene is that, here, the BdG for\-ma\-lism includes both spins for electrons and holes. Therefore, when adopting previous results, we are required to pro\-per\-ly account for a factor of $\nicefrac{1}{2}$ so not to double count the electronic degrees of freedom. See also Appendix~\ref{app:AppendixStiffness}. Under these conditions, we find the following result:
\begin{align}
D=\frac{1}{4}\frac{\pi E_F}{\Phi_0^2}\left[f_\delta-\delta^2\ln\left(\frac{\delta}{1+f_\delta}\right)\right].
\end{align}

\subsection{Rashba Metal}\label{sec:Rashba}

We now repeat the above procedure for a Rashba metal. In contrast to the TI case, here, the Fermi ener\-gy is the dominant energy scale, i.e., much larger than the Rashba splitting and the pairing gap. In fact, for certain calculations within this so-called quasiclassical limit, $E_F$ can be taken to be equal to positive infinity. In this sense, a number of the physical quantities that we aim at eva\-lua\-ting depend only on $\delta=\Delta/E_{\rm soc}$, where once again the Rashba SOC energy is defined as $E_{\rm soc}=\upsilon_R\hbar k_F$, but with the Fermi wave number being set by the Schr\"odinger kinetic energy to the value $k_F=\sqrt{2mE_F}/\hbar$.

The expression for the magnetoelectric coefficient stemming from broken inversion was previously obtained in Refs.~\onlinecite{PershogubaCurrents} and~\onlinecite{Anatomy} in different parameter regimes. In the regime of interest here, we adopt the result of Ref.~\onlinecite{Anatomy}, which takes the following form:
\begin{align}
\chi_R=-\frac{1}{4\Phi_0}\left[1+\left(\frac{\delta}{f_\delta}\right)^2+\frac{\delta^2}{f_\delta}\left(2+\frac{1}{f_\delta^2}\right)\ln\left(\frac{\delta}{1+f_\delta}\right)\right],
\end{align}

\noi and has been obtained for a disorder-free system.

The out-of-plane spin susceptibility was also calculated in Ref.~\onlinecite{Anatomy} under the same assumptions, and takes the following form:
\begin{align}
\chi_\perp^{\rm spin}({\rm v})=\frac{m}{\pi \hbar^2}\left[1-\frac{\ln\big({\rm v}+f_{{\rm v}}\big)}{{\rm v}f_{{\rm v}}}\right],\label{eq:SpinSuscMetal}
\end{align}

\noi where the prefactor of the bracket corresponds to the
spin-summed normal phase density of states evaluated
at the Fermi level when the Rashba SOC is discarded. In the above, we chose to parametrize the spin su\-sceptibility more compactly in terms of the inverse of $\delta$, i.e., ${\rm v}=E_{\rm soc}/\Delta$.
With the above result at hand, we now obtain the Zeeman contribution $\chi_Z$. For a system with $g=2$ and $m=m_e$, we find the following expression:
\begin{align}
\chi_Z({\rm v})=\frac{1}{2\Phi_0}\left[1-\frac{\ln\big({\rm v}+f_{{\rm v}}\big)}{{\rm v}f_{{\rm v}}}\right].
\end{align}

Next in line is the magnetic stiffness $c_M$. Following the procedure described in Appendix~\ref{app:AppendixMagStiff}, we find that at lea\-ding order in the dimensionless parameters $u=\Delta/E_F$ and $u{\rm v}=E_{\rm soc}/E_F$ the stiffness takes the form:\begin{align}
c_M=\frac{1}{16\pi E_F}\left\{1+\frac{{\rm v}^2}{2}+\frac{1}{3}\frac{16}{(u{\rm v}^2)^2}-\frac{2}{3}\left[u^3\left(\frac{{\rm v}}{2}\right)^5\right]^2\right\}\,.
\label{eq:cMRashba}
\end{align}

\noi The above expression is obtained by adopting the quasiclassical framework, within which the helical dispersions $\xi_\pm(k)=(\hbar k)^2/2m\pm\upsilon_R\hbar k$ are linearized according to $\xi_\pm(k)=\hbar\upsilon_F(k-k_{F_\pm})$, where we introduced the Fermi wave numbers $k_{F_\pm}=\big(1\mp\upsilon_R/\upsilon_F\big)k_F$ along with the Fermi velocity which is given as usual by $\upsilon_F=\hbar k_F/m_e$.

The next quantity to be determined is the superfluid stiffness of a superconducting Rashba metal. While the superfluid stiffness of a one-dimensional metallic SC with a large Fermi ener\-gy has been previously obtained in Ref.~\onlinecite{WangAssiliPKarticle}, to our knowledge the stiffness for a superconducting Rashba metal in two spatial dimensions remains unaddressed. Here, we also follow the adiabatic approach introduced in Refs.~\onlinecite{WangAssiliPKletter,WangAssiliPKarticle}, and after employing the quasiclassical approximation we  obtain the expression:
\begin{align}
D=\frac{\pi E_F}{\Phi_0^2}\,.
\end{align}

\section{Out-of-plane Spin Susceptibility}\label{app:AppendixSusce}

In this paragraph, we provide further details concer\-ning the evaluation of the spatially-uniform out-of-plane spin susceptibility $\chi_\perp^{\rm spin}$. We restrict our study to the case of superconducting TI states, since the quantity of interest was previously obtained in Ref.~\onlinecite{Anatomy} for a supercon\-duc\-ting Rashba metal. As it is was already pointed out in that prior work, the most convenient way to obtain $\chi_\perp^{\rm spin}$ is by including a uniform out-of-plane magnetization $M_z$ to the Hamiltonian of the Rashba SC. The respective susceptibility is then obtained from the defining relation:
\begin{align}
\chi_\perp^{\rm spin}=-\left.\frac{d^2E_{\rm gs}}{dM_z^2}\right|_{M_z=0}\,,
\end{align}

\noi where $E_{\rm gs}$ is the ground state energy of the system per area for a nonzero $M_z$. In the additional presence of the magnetization, the resulting wave vector space Hamiltonian in the limit $m\rightarrow\infty$, which becomes relevant when discussing the TI system, takes the form:
\begin{align}
\hat{\cal H}(\bm{k})=\tau_z\big[\upsilon_R\hbar(k_x\sigma_y-k_y\sigma_x)-\mu\big]+\Delta\tau_x-M_z\sigma_z\,,
\end{align}

\noi and gives rise to the eigenenergies $\pm E_\pm(\omega)$, where we set:
\begin{align}
E_\pm(\omega)=\sqrt{\omega^2+\mu^2+M_z^2+\Delta^2\pm2\sqrt{{\cal R}(\omega)}}\,.
\end{align}

\noi In the above, we simplified the notation by employing the variable $\omega=\upsilon_R\hbar k$, where $k=|\bm{k}|$. In addition, we defined the function:
\begin{align}
{\cal R}(\omega)=\mu^2\big(\omega^2+M_z^2\big)+\big(M_z\Delta\big)^2\,.
\end{align}

\noi Since the spectrum depends only on $k$, and thus only on $\omega$, we carry out the trivial integration over the angular coordinate and find that the ground state energy per area is given by the expression:
\begin{align}
E_{\rm gs}=-\frac{1}{2}\sum_{s=\pm}\int_0^\Lambda\frac{d\omega}{2\pi(\upsilon_R\hbar)^2}\,\omega E_s(\omega)\,,
\end{align}

\noi where we introduced the ``ultraviolet" energy cutoff $\Lambda$. Note that the factor of $\nicefrac{1}{2}$ enters to ensure that the electronic degrees of freedom are correctly counted within the here-employed BdG formalism. The above integration is
straightforward and the result depends strongly on $\Lambda$. In
Ref.~\onlinecite{Anatomy}, the contribution originating from higher energies was discarded. After additional numerical checks, we here conclude that the high-energy contribution is re\-le\-vant and the response is paramagnetic. By taking
the limit $\Lambda\gg E_F,\Delta$ we recover the result shown in Eq.~\eqref{eq:SpinSusc}.

\section{Magnetic Stiffness}
\label{app:AppendixMagStiff}

In this section, we demonstrate the method to eva\-lua\-te the magnetic stiffness $c_M$. We start by the spatially-integrated energy in the presence of a spatially-varying magnetization $M_z(\bm{r})=\int d\bm{q}\, e^{i\bm{q}\cdot\bm{r}}M_z(\bm{q})/(2\pi)^2$. At se\-cond order in the Fourier components $M_z(\bm{q})$ of the magnetization, the integrated energy reads as:
\begin{align}
E_{M_z^2}=-\frac{1}{2}
\int\frac{d\bm{q}}{(2\pi)^2}\, M_z(-\bm{q})\chi_\perp^{\rm spin}(\bm{q})M_z(\bm{q})\,,
\end{align}

\noi where we have introduced the out-of-plane spin susceptibility defined for a nonzero wave vector transfer $\bm{q}$:
\begin{align}
\chi_\perp^{\rm spin}(\bm{q})=-\frac{1}{2}\int dK\,{\rm Tr}\left[\sigma_z\hat{\cal G}_0(\bm{k}+\bm{q},\epsilon)\sigma_z\hat{\cal G}_0(\bm{k},\epsilon)\right].
\end{align}

\noi In the above, we made use of the shorthand notation:\begin{align}
\int dK=\int\frac{d\bm{k}}{(2\pi)^2}\int_{-\infty}^{+\infty}\frac{d\epsilon}{2\pi}\,,
\end{align}

\noi and introduced the matrix Green function:
\begin{align}
\hat{{\cal G}}_0^{-1}(\epsilon,\bm{k})=i\epsilon-\hat{{\cal H}}_0(\bm{k})\,.\label{eq:Green}
\end{align}

On the other hand, the respective spatially-integrated term appearing in our GL functional, reads as:
\begin{align}
E_{{\rm GL}-M_z^2}=\frac{c_M}{2}
\int\frac{d\bm{q}}{(2\pi)^2}\, M_z(-\bm{q})q^2M_z(\bm{q})\,,
\end{align}

\noi where we have $q=|\bm{q}|$. By identifying the two expressions for the integrated energy, we find the defining formula:
\begin{align}
c_M=-\frac{1}{4}\left[\frac{\partial_{q_x}^2\chi_\perp^{\rm spin}(\bm{q})}{\partial q_x^2}+\frac{\partial_{q_y}^2\chi_\perp^{\rm spin}(\bm{q})}{\partial q_y^2}\right]_{\bm{q}=\bm{0}}\,.
\end{align}

\noi Using the above expression, we obtain the results in Eqs.~\eqref{eq:cMTI} and~\eqref{eq:cMRashba}.

\section{Superfluid Stiffness}
\label{app:AppendixStiffness}

To obtain the superfluid stiffness, we suitably adopt the adiabatic approach presented in Refs.~\onlinecite{WangAssiliPKletter,WangAssiliPKarticle}. In contrast to the expressions obtained in these prior works, here, we have to be cautious and account for a factor of $\nicefrac{1}{2}$ in order to avoid double-counting the degrees of freedom within the BdG description. Hence, the elements $D_{ij}$ of the superfluid stiffness tensor with $i,j=x,y$, read as:
\begin{align}
D_{ij}=\left(\frac{\pi}{\Phi_0}\right)^2\int dK\ph{\rm Tr}\Big\{\big[\partial_{k_i}\hat{h}(\bm{k})\big]\mathds{1}_\tau\hat{\cal F}_{k_j\phi}(\epsilon,\bm{k},\phi)\Big\},
\label{eq:StartingPoint}
\end{align}

\noi where we introduced the normal phase Hamiltonian:
\begin{align}
\hat{h}(\bm{k})=\frac{(\hbar \bm{k})^2}{2m}-\mu+\upsilon_R\hbar\big(k_x\sigma_y-k_y\sigma_x\big)\,,
\end{align}

\noi along with the matrix function $\hat{\cal F}_{k_j\phi}(\epsilon,\bm{k},\phi)$, defined as:
\begin{align}
\hat{\cal F}_{k_j\phi}=\nicefrac{1}{2}\big(\partial_\epsilon\hat{{\cal G}}^{-1}\big)\hat{{\cal G}}\big(\partial_\phi\hat{{\cal G}}^{-1}\big)\hat{{\cal G}}\big(\partial_{k_j}\hat{{\cal G}}^{-1}\big)\hat{{\cal G}}-\partial_\phi\leftrightarrow\partial_{k_j}.
\end{align}

\noi In the above, we suppressed the arguments of the va\-rious functions for notational convenience and, most importantly, we introduced the matrix Green function through:
\begin{align}
\hat{{\cal G}}^{-1}(\epsilon,\bm{k},\phi)=i\epsilon-\hat{{\cal H}}(\bm{k},\phi)\,,\label{eq:SynthGreen}
\end{align}

\noi which is defined in the synthetic energy-wave vector-phase space and results from the adiabatic Hamiltonian:
\begin{align}
\hat{{\cal H}}(\bm{k},\phi)=\hat{h}(\bm{k})\tau_z+\Delta\tau_xe^{-i\phi\tau_z}\label{eq:SynthHam}
\end{align}

\noi that is also defined in wave vector-phase $(\bm{k},\phi)$ space. By virtue of the rotational symmetry of the model which yields $D_{xx,yy}=D$, we equivalently obtain the superfluid stiffness from the expression $D=(D_{xx}+D_{yy})/2$. By employing the above framework, we recover the expression for the superfluid stiffness discussed in Appendix~\ref{app:AppendixGLcoeff}.

\section{Vortex-MZMs on TI Surfaces}\label{app:AppendixMZM}

In the following paragraphs, we provide additional details regar\-ding the emergence of domain-wall-MZMs in the case of a zero-field superconducting vortex induced by a magnetic island. For our analysis, we consider a simplified version of the continuum model for the superconducting TI helical surface states described in Eq.~\eqref{eq:BdGHamiltonian}. Under the same assumptions discussed in the main text, as well as after setting $E_F=0$ and taking the limit $m\rightarrow\infty$ with no loss of generality, we go beyond the adia\-batic framework presented in the main text and end up with the following BdG Hamiltonian for a cylindrically-symmetric spatial configuration of the va\-rious fields:
\begin{align}
\hat{{\cal H}}(\rho,\theta)=\upsilon_R\tau_z\big(\hat{\bm{p}}\times\bm{\sigma}\big)\cdot\hat{\bm{z}}-\tilde{I}_z(\rho)\sigma_z+\Delta(\rho)e^{i\nu_\phi\theta\tau_z}\tau_x,\label{eq:Hamiltonian4Vortex}
\end{align}

\noi where $\nu_\phi\in\mathbb{Z}$. Here, $\Delta(\rho)$ is considered to be zero at least at the origin of the coordinate system $\rho=0$, since the superconducting coherence length $\xi_S$ is the smallest lengthscale in the problem of interest. Expres\-sing $\hat{\bm{p}}$ using the cylindrical coordinates $(\rho,\theta)$ yields the Hamiltonian:
\bea
\hat{{\cal H}}_{\rm BdG}(\rho,\theta)=\hat{\cal U}_{\nu_\phi}(\theta)\hat{{\cal H}}_{\rm BdG}'(\rho,\theta)\,\hat{\cal U}_{\nu_\phi}^{\dag}(\theta)\,,
\eea

\noi where we defined:
\bea
\hat{{\cal H}}_{\rm BdG}'(\rho,\theta)&=&\upsilon_R\tau_z\sigma_y\bigg[
\hat{p}_\rho+\frac{\hbar(1-{\nu_\phi}\tau_z\sigma_z)}{2\rho i}\bigg]
\no\\
&-&\frac{\upsilon_R\hat{L}_z}{\rho}\tau_z\sigma_x-\tilde{I}(\rho)\sigma_z+\Delta(\rho)\tau_x,\quad
\label{eq:TIMZMequationEarly}
\eea

\noi with $\hat{L}_z=-i\hbar\partial_\theta$ denoting the out-of-plane orbital angular momentum operator. In the above, we employed a unitary transformation effected by the unitary operator $
\hat{\cal U}_{\nu_\phi}(\theta)={\rm Exp}\big[i\theta\tau_z(\nu_\phi-\tau_z\sigma_z)/2\big]$. We proceed by choo\-sing a basis that absorbs $\hat{\cal U}_{\nu_\phi}(\theta)$, so that the Hamiltonian is given only by $\hat{{\cal H}}_{\rm BdG}'(\rho,\theta)$. Since $\hat{\cal U}_{\nu_\phi}(\theta)$ depends on the vorticity, the transition to the new basis affects the energy spectrum through a periodicity constraint that it imposes on the eigensolutions $\bm{\Phi}'(\rho,\theta)$ of the BdG Hamiltonian, which are defined through $\hat{{\cal H}}_{\rm BdG}'(\rho,\theta)\bm{\Phi}'(\rho,\theta)=E\bm{\Phi}'(\rho,\theta)$. Specifically, eve\-ry $\bm{\Phi}'(\rho,\theta)$ solution is required to satisfy $
\bm{\Phi}'(\rho,\theta+2\pi)=\mp\bm{\Phi}'(\rho,\theta)$, with the $-$ ($+$) corresponding to $\nu_\phi\in2\mathbb{Z}$ $\big(\nu_\phi\in2\mathbb{Z}+1\big)$. See also Refs.~\cite{ReadGreen,Volovik99,Alicea}. By taking advantage of the fact that $\hat{{\cal H}}_{\rm BdG}'(\rho,\theta)$ depends on $\theta$ only in terms of the operator $\hat{L}_z$, we employ the expansion $\bm{\Phi}'(\rho,\theta)=\sum_ne^{in\theta}\bm{\Phi}_n'(\rho)$. After the above, the periodicity constraint implies that $n\in\mathbb{Z}+1/2$ for $\nu_\phi\in2\mathbb{Z}$, while  $n\in\mathbb{Z}$ for $\nu_\phi\in2\mathbb{Z}+1$. Lastly, since the MZM eigenvector $\bm{\Phi}_0'(\rho)$ corresponds to ener\-gy $E=0$ and $n=0$, the MZM becomes accessible only for $\nu_\phi\in2\mathbb{Z}+1$, and is determined by the equation:
\begin{align}
\bigg(\frac{d}{d\rho}+\frac{{\cal P}_{\nu_\phi}}{\rho}\bigg)\bm{\Phi}_0'(\rho)=
\frac{\Delta(\rho)\tau_y\sigma_y-\tilde{I}_z(\rho)\tau_z\sigma_x}{\upsilon_R\hbar}\ph\bm{\Phi}_0'(\rho)\,,
\label{eq:TIMZMequation}
\end{align}

\noi where we defined ${\cal P}_{\nu_\phi}=(1-\nu_\phi\tau_z\sigma_z)/2$ with $\nu_\phi\in2\mathbb{Z}+1$.

\subsection{Fu-Kane Model}

We proceed by first briefly reminding the reader of the emergence of MZMs in the Fu-Kane model~\cite{FuKane}. For convenience, we assume for the rest of this paragraph that $|\nu_\phi|=1$, in which case ${\cal P}_{\nu_\phi}$ becomes a pro\-jec\-tor ope\-ra\-tor. Since in the Fu-Kane model $\tilde{I}_z(\rho)$ is zero and as a result $[{\cal P}_{\nu_\phi},\tau_y\sigma_y]=0$, the two different subspaces spanned by the projectors ${\cal P}_{\pm1}$ become decoupled. In fact, the MZM solution comes only from one of the two subsectors. Since the MZM solution needs to be properly regularized for $\rho\rightarrow0$, the only acceptable solution constitutes the eigenstate of $\tau_y\sigma_y$ with eigenvalue $-{\rm sgn}(\upsilon_R)$ which addi\-tio\-nal\-ly yields ${\cal P}_{\nu_\phi}=0$. The respective spatial part of the MZM wavefunction $\Phi_0'(\rho)$ reads as:
\begin{align}
\Phi_0'(\rho)=\Phi_0'(\rho=0){\rm Exp}\left[-\int_0^\rho d\bar{\rho}\ph\frac{\Delta(\bar{\rho})}{|\upsilon_R|\hbar}\right].
\end{align}
We remark that the emergence of the MZM within the Fu-Kane model~\cite{FuKane} immediately follows from the presence of the band touching point in the energy spectrum of the helical surface states. This can be either understood through the connection~\cite{Chamon} of the Fu-Kane model to the Jackiw-Rossi model~\cite{JRossi}, or, by analyzing the topological invariant dictating such a defect configuration~\cite{SteffensenMZM}. Alternatively, the MZM can be viewed as the zero mode appearing due to the effective energy spectrum gap clo\-sing which takes place in the region where $\Delta(\rho)=0$.

\subsection{Our Model}

In stark contrast to the Fu-Kane framework, a different mechanism becomes relevant in our work. The addition of $\tilde{I}_z(\rho)$ implies that the presence of the vortex cannot effectively close the energy gap at the points of space where $\Delta(\rho)=0$. The MZM appears here as a consequence of a Majorana chiral edge mode which is trapped at the domain wall where $\tilde{I}_z(\rho)-|\Delta(\rho)|$ changes sign, say at $\rho=\rho_{\rm dw}$. For a vorticity value $\nu_\phi\in2\mathbb{Z}+1$ ($\nu_\phi\in2\mathbb{Z}$), the Majorana chiral edge mode is dictated in low energies by the dispersion relation $E_n=nE_0$, where $n\in\mathbb{Z}$ ($n\in\mathbb{Z}+1/2$). Note that the characteristic energy scale $E_0$ is approximately $E_0\approx\hbar\upsilon_R/\rho_{\rm dw}$ for a sufficiently large $\rho_{\rm dw}$ that allows us to ignore the effects of curvature, see for instance the analysis in Ref.~\onlinecite{Alicea}. Hence, the pre\-sen\-ce of the vortex is required here only to ensure that the Majorana chiral edge mode dispersion crosses the zero energy axis. The domain-wall-MZM is identified with the $n=0$ mode which is uniformly delocalized along the $\rho=\rho_{\rm dw}$ orbit. Notably, $\rho_{\rm dw}$ can still be quite small and comparable to $\xi_S$, so that the MZM spectral weight may appear in experiments to be concentrated near $\rho=0$.

A precise analytical solution is obtainable also in the present case. We start from Eq.~\eqref{eq:TIMZMequation}, we reorder the various terms, and define $\bm{\Phi}_0'(\rho)=\bm{F}(\rho)/\sqrt{\rho}$. These steps yield the following equation for the auxiliary MZM state vector $\bm{F}(\rho)$:
\begin{align}
\frac{d\bm{F}(\rho)}{d\rho}=\left[
\frac{\nu_\phi\tau_z\sigma_z}{2\rho}+\frac{\Delta(\rho)\tau_y\sigma_y-\tilde{I}_z(\rho)\tau_z\sigma_x}{\upsilon_R\hbar}\right]\bm{F}(\rho)\,.
\label{eq:TIMZMequationAux}
\end{align}

\noi The above expression immediately provides the four possible outcomes for $\bm{F}(\rho)$, since the r.h.s. can be readily diagonalized by employing eigenstates of the charge conjugation operator $\Xi=\tau_y\sigma_y{\cal K}$, where ${\cal K}$ corresponds to the complex conjugation operator. We thus find the four possibilities for the initial MZM state vector:
\begin{align}
\bm{\Phi}_0'(\rho)\propto\frac{\bm{\Phi}_0'(\rho=0)}{\sqrt{\rho}}{\rm Exp}\left[\mp\int_0^\rho d\bar{\rho}\ph
\frac{\Delta(\bar{\rho})\mp m(\bar{\rho})}{|\upsilon_R|\hbar}\right],
\end{align}

\noi where in the above we introduced the positive function $
m(\rho)=\sqrt{\tilde{I}_z^2(\rho)+\big[\nu_\phi\upsilon_R\hbar/(2\rho)\big]^2}$. By virtue of the fact that $\lim_{\rho\rightarrow\infty}m(\rho)=0$, a normalizable solution for $\rho\rightarrow0$ and $\rho\rightarrow\infty$ can be only obtained from the eigenstates which have an exponent: $-\int_0^\rho d\bar{\rho}\ph
\big[\Delta(\bar{\rho})-m(\bar{\rho})\big]/(|\upsilon_R|\hbar)$.

One observes that the above solution smoothly tends to the one of the Fu-Kane model as $\tilde{I}_z(\rho)\rightarrow0$. For a nonzero $\tilde{I}_z(\rho)$ the final result depends on the precise expressions for the profiles of $\Delta(\rho)$ and $\tilde{I}_z(\rho)$. As a concrete example, we consider that $\Delta(\rho)=\Delta\Theta(\rho-\xi_S)$ and $\tilde{I}_z(\rho)=2\tilde{I}_z\Theta(\tilde{\rho}_I-\rho)$~\cite{CommentOnDiskProfile}, where $\tilde{\rho}_I$ is the effective radius of the magnetic island which is dictated by uniform out-of-plane magnetization. On the other hand, $\xi_S$ determines the radius of the superconducting vortex core.

The above conclusions can be alternatively obtained by means of projec\-ting Eq.~\eqref{eq:TIMZMequation} onto the rim MZM solution which is obtained for $n=0$, and is identified with the only one properly-regularized eigensolution out of the two eigenstates of the operator $\Delta(\rho)\tau_x-\tilde{I}_z(\rho)\sigma_z$ with eigenvalues $\pm\big[\Delta(\rho)-\tilde{I}_z(\rho)\big]$. After the projection, the term $\propto{\cal P}_{\nu_\phi}$ drops out from Eq.~\eqref{eq:TIMZMequation}, and the emergence of the domain-wall-MZM maps to the scenario predicted for a TI edge~\cite{FuKane,KotetesTI,Pekker}, where a MZM is trapped at a domain wall where the magnetic and pairing gaps compensate each other. However, there is a crucial difference here. For the TI edge the MZM is always accessible as long as such a domain wall is established. In contrast, here, the MZM arises only for a vortex in the phase field of the pairing gap which carries an odd value of vorticity.


\begin{thebibliography}{00}
\bibitem{CommentOnYSR} The eigenvalue problem of a Rashba SC in the presence of a magnetic island reads as $\big[\hat{h}_{\rm sc}(\hat{\bm{p}})-I_z(\bm{r})\sigma_z\big]\bm{\chi}(\bm{r})=\varepsilon\bm{\chi}(\bm{r})$. Here, $\bm{\chi}(\bm{r})$ is the wavefunction and $\hat{h}_{\rm sc}(\hat{\bm{p}})$ is the translationally invariant bare Hamiltonian of the SC. We transfer to momentum space~\cite{Heimes} and find $\bm{\chi}(\bm{p})=- \hat{g}_{\rm sc}(\varepsilon,\bm{p})\int \,d\bm{q}\,f(\bm{p}-\bm{q})S_z\sigma_z\bm{\chi}(\bm{q})/(2\pi)^2$, where we introduced the Green function $\hat{g}_{\rm sc}^{-1}(\varepsilon,\bm{p})=\varepsilon-\hat{h}_{\rm sc}(\bm{p})$, and defined $f(\bm{q})=1/\big[1+(q\rho_I)^2\big]$ after assu\-ming the profile in Eq.~\eqref{eq:SpinMomentProfile}. For a magnetic impurity we have $\rho_I\rightarrow0$ and the above equation becomes $\bm{\chi}(\bm{r})=-\hat{g}_{\rm sc}(\varepsilon,\bm{r})S_z\sigma_z\bm{\chi}(\bm{r}=\bm{0})$, with the Fourier transform $\hat{g}_{\rm sc}(\varepsilon,\bm{r})=\int d\bm{p}\,e^{i\bm{p}\cdot\bm{r}}\hat{g}_{\rm sc}(\varepsilon,\bm{p})/(2\pi)^2$. This is precisely the YSR regime with YSR eigen\-ener\-gies obtained by the eigenvalue problem for $\bm{r}=\bm{0}$, i.e., $\hat{g}_{\rm sc}(\varepsilon,\bm{r})+S_z^{-1}\sigma_z=0$. For a magnetic island $\rho_I$ has to be compared to the modulus $|\bm{p}-\bm{q}|$ of the momentum transfer vector $\bm{p}-\bm{q}$ connec\-ting $\bm{\chi}(\bm{p})$ and $\bm{\chi}(\bm{q})$. For a large Fermi energy, $\bm{p}$ and $\bm{q}$ both belong to the Fermi surface, hence $|\bm{p}-\bm{q}|$ is close to zero ($2k_F$) for the so-called forward (backward) scattering scenario. When $2k_F\rho_I\gg1\rightarrow4\pi\rho_I\gg\lambda_F$ only forward scattering becomes relevant and $\bm{p}\approx\bm{q}$. Hence, the island does not lead to YSR states but only affects the dispersive bulk states, since the electrons cannot get backscattered and trapped inside the magnetic island. Thus, here it is eligible to restrict to bulk physics.
\bibitem{CommentOnZeeman} This is understood by recalling that the original expression of the Zeeman term within the BdG framework takes the form $E_Z=gi\big(\hat{\bm{\pi}}\times\hat{\bm{\pi}}\big)\cdot\tau_z\bm{\sigma}/(4m)$. Therefore, direct replacement of $\hat{\bm{\pi}}$ yields the Zeeman term in Eq.~\eqref{eq:BdGHamiltonian}. When we instead consider a gauge in which $\hat{\bm{\pi}}$ does not include the gradient of the superconducting phase, one retrieves the standard expression $E_Z=g\mu_B\bm{B}\cdot\bm{\sigma}/2$.
\bibitem{CommentOnSymmetrization} This symmetrization allows us to avoid any ambiguity in the choice of the form of the coupling, since any linear combination $c_1I_z(\bm{r})\hat{\bm{z}}\cdot\big[\bm{\nabla}\times\bm{a}(\bm{r})\big]+c_2\bm{a}(\bm{r})\cdot\big[\bm{\nabla}\times\hat{\bm{z}}I_z(\bm{r})\big]$, with $c_1+c_2={\cal X}$ does not alter the bulk physics. Note that either term leads to a legitimate form for the coupling. This is because the coordinate space form of the coupling is obtained from effecting a Fourier transform on a momentum space result which is evaluated for an infinite system with no boundaries. See for instance Ref.~\onlinecite{Anatomy} for additional technical details. This momentum space result can be expressed into two equi\-va\-lent forms, i.e., $\int d\bm{q}\, e^{i\bm{q}\cdot\bm{r}}I_z(-\bm{q})\hat{\bm{z}}\cdot\big[i\bm{q}\times\bm{a}(\bm{q})\big]$ and $\int d\bm{q}\, e^{i\bm{q}\cdot\bm{r}}\bm{a}(-\bm{q})\cdot\big[ i\bm{q}\times\hat{\bm{z}}I_z(\bm{q})\big]$. Hence, by symmetri\-zing the bulk result no ambiguous surface terms emerge.
\bibitem{CommentOnSuc} Note that in Ref.~\onlinecite{Anatomy}, the coefficient $\chi_R$ was denoted $\chi$.
\bibitem{CommentOnDiskProfile} In the case of a disk-like spatial profile for the spin-moment field $I_z(\bm{r})=I_{\rm disc}\Theta(\rho_I-\rho)$ with $\Theta$ deno\-ting the Heaviside unit-step function, we define the coefficient through $S_z=\int d\bm{r}\, I_z(\bm{r})$, and find $I_{\rm disc}=S_z/\big(\pi\rho_I^2\big)$. Hence the present coefficient only differs by a factor of $2$ compared to the one obtained for the profile which is proportional to $K_0(\rho/\rho_I)$. Here, the Hankel transform yields $I_z(\bm{q})=2S_zJ_1(q\rho_I)/(q\rho_I)$, where $J_1(z)$ is the first order Bessel function of the first kind with $z\in[0,\infty)$.
\bibitem{CommentOnDecomposition} Starting from Eq.~\eqref{eq:Evortex}, we write that the energy of a single vortex with vorticity $\nu$ is expressed using the equi\-va\-lent but more convenient form $E_\nu=\nu^2/2-S\nu$, with $S$ the source term. We explore whether it is energetically more favorable to decompose a single vortex with vorticity $\nu$ into two vortices with vorticities $\nu_{1,2}$ where $\nu_1+\nu_2=\nu$. The energy of two such vorticies reads as $E_{\nu_{1,2}}=\nu_1^2/2-S\nu_1+\nu_2^2/2-S\nu_2\equiv \nu_1^2/2+\nu_2^2/2-S\nu$. With no loss of generality, we make the replacement $\nu_2=\nu-\nu_1$ and find that $E_{\nu_{1,2}}-E_\nu=\nu_1(\nu_1-\nu)$. The term on the r.h.s. is minimized for the integer value of $\nu_1$ closest to $\nu/2$. The value of this term at this minimum is negative, thus, implying that higher win\-ding vortices are unstable, precisely similar to Abrikosov vortices~\cite{deGennes}.
\bibitem{CommentSmallFermiEnergyRashba} Note that in the case of a small Fermi energy which is comparable to $\Delta$, the analysis that was carried out in Sec.~\ref{sec:Rashba} is no longer valid. In this limit, the va\-rious parameters which govern the stabilization of superconducting vortices can be obtained by extending the results of Sec.~\ref{sec:TIstates} to the case of two helical branches.
\end{thebibliography}
\end{document}